\documentclass[notitlepage,a4paper,aps,prd,onecolumn,superscriptaddress,nofootinbib,groupedaddress,longbibliography]{revtex4-1}
\usepackage{amsmath}
\usepackage{amsfonts}
\usepackage{amssymb}
\usepackage{graphicx}
\usepackage[utf8x]{inputenc}
\usepackage[normalem]{ulem}
\usepackage[T1]{fontenc}
\usepackage[mathscr]{euscript}
\usepackage[svgnames]{xcolor}
\usepackage{tikz}
\usepackage{times}
\usepackage{verbatim}
\usetikzlibrary{arrows,shapes}
\usetikzlibrary{matrix}
\usetikzlibrary{shadows}
\usetikzlibrary{positioning}
\usepackage{amsbsy,amscd,latexsym}

\usepackage{subfigure}
\usepackage{epsfig}
\usepackage{graphicx}
\usepackage[all,knot]{xy}\xyoption{arc}
\usepackage{url}
\usepackage[colorlinks]{hyperref}
\usepackage{setspace}
\graphicspath{{images/}}
\usepackage[vlined]{algorithm2e}
\usepackage{time}
\usepackage{relsize}
\usepackage{accents}
\usepackage{multirow}
\usepackage{palatino}
\usepackage{ae}

\newcommand{\LG}{\mathcal{L}}
\newcommand{\HG}{\mathcal{H}}
\newcommand{\inducee}{\accentset{\mathfrak{e}}{\gamma}}
\newcommand{\inducef}{\accentset{\mathfrak{f}}{\gamma}}
\newcommand{\induceh}{\accentset{\text{\ae}}{\gamma}}
\newcommand{\lapsee}{\accentset{\mathfrak{e}}{N}}
\newcommand{\lapsef}{\accentset{\mathfrak{f}}{N}}
\newcommand{\lapseh}{\accentset{\text{\ae}}{N}}
\newcommand{\shifte}{\accentset{\mathfrak{e}}{N}}
\newcommand{\shiftf}{\accentset{\mathfrak{f}}{N}}
\newcommand{\shifth}{\accentset{\text{\ae}}{N}}

\newcommand{\nn}{\nonumber}

\newcommand{\pd}{\partial}
\newcommand{\td}{\mathrm{d}}

\newcommand{\req}{\stackrel{!}{=}}

\newcommand{\modu}[1]{\left\vert{#1}\right\vert}

\newcommand{\gbe}{\accentset{\mathfrak{e}}{\overline{g}}}

\newcommand{\ee}{\mathfrak{e}}
\newcommand{\ff}{\mathfrak{f}}
\newcommand{\gee}{\accentset{\mathfrak{e}}{g}}
\newcommand{\gf}{\accentset{\mathfrak{f}}{g}}
\newcommand{\gh}{\accentset{\text{\ae}}{g}}
\newcommand{\he}{\accentset{\mathfrak{e}}{h}}
\newcommand{\hf}{\accentset{\mathfrak{f}}{h}}
\newcommand{\be}{\accentset{\mathfrak{e}}{b}}
\newcommand{\bff}{\accentset{\mathfrak{f}}{b}}
\newcommand{\Be}{\accentset{\mathfrak{e}}{B}}

\newcommand{\normee}{\accentset{\mathfrak{e}}{\xi}}
\newcommand{\normff}{\accentset{\mathfrak{f}}{\xi}}
\newcommand{\normhh}{\accentset{\text{\ae}}{\xi}}
\newcommand{\spatialee}{{}^3\overset{\mathfrak{e}}{\mathbb{T}}}
\newcommand{\spatialff}{{}^3\overset{\mathfrak{f}}{\mathbb{T}}}
\newcommand{\spatialhh}{{}^3\overset{\text{\ae}}{\mathbb{T}}}
\newcommand{\torsionee}{\accentset{\mathfrak{e}}{T}}
\newcommand{\torsionff}{\accentset{\mathfrak{f}}{T}}
\newcommand{\CeVec}{{}^{\mathcal{V}}\accentset{\mathfrak{e}}{C}}
\newcommand{\CeAn}{{}^{\mathcal{A}}\accentset{\mathfrak{e}}{C}}
\newcommand{\CeSym}{{}^{\mathcal{S}}\accentset{\mathfrak{e}}{C}}
\newcommand{\CeTr}{{}^{\mathcal{T}}\accentset{\mathfrak{e}}{C}}
\newcommand{\CfVec}{{}^{\mathcal{V}}\accentset{\mathfrak{f}}{C}}
\newcommand{\CfAn}{{}^{\mathcal{A}}\accentset{\mathfrak{f}}{C}}
\newcommand{\CfSym}{{}^{\mathcal{S}}\accentset{\mathfrak{f}}{C}}
\newcommand{\CfTr}{{}^{\mathcal{T}}\accentset{\mathfrak{f}}{C}}
\newcommand{\CkineVec}{{}^{\mathcal{V}}\accentset{\mathfrak{e}}{\mathcal{C}}}
\newcommand{\CkineAn}{{}^{\mathcal{A}}\accentset{\mathfrak{e}}{\mathcal{C}}}
\newcommand{\CkineSym}{{}^{\mathcal{S}}\accentset{\mathfrak{e}}{\mathcal{C}}}
\newcommand{\CkineTr}{{}^{\mathcal{T}}\accentset{\mathfrak{e}}{\mathcal{C}}}
\newcommand{\CkinfVec}{{}^{\mathcal{V}}\accentset{\mathfrak{f}}{\mathcal{C}}}
\newcommand{\CkinfAn}{{}^{\mathcal{A}}\accentset{\mathfrak{f}}{\mathcal{C}}}
\newcommand{\CkinfSym}{{}^{\mathcal{S}}\accentset{\mathfrak{f}}{\mathcal{C}}}
\newcommand{\CkinfTr}{{}^{\mathcal{T}}{\accentset{\mathfrak{f}}{\mathcal{C}}}}

\newcommand{\CHRVecg}{{}^\mathcal{V}\accentset{\mathfrak{e}}{\mathfrak{C}}}
\newcommand{\CHRAng}{{}^\mathcal{A}\accentset{\mathfrak{e}}{\mathfrak{C}}}
\newcommand{\CHRg}{\accentset{\mathfrak{e}}{\mathfrak{C}}}

\newcommand{\CHRf}{\accentset{\mathfrak{f}}{\mathfrak{C}}}
\newcommand{\CHRmg}{{}^\mathcal{M}\accentset{\mathfrak{e}}{\mathfrak{C}}}
\newcommand{\CHRmf}{{}^\mathcal{M}\accentset{\mathfrak{f}}{\mathfrak{C}}}
\newcommand{\conje}{\accentset{\mathfrak{e}}{\pi}}
\newcommand{\conjf}{\accentset{\ff}{\pi}}
\newcommand{\conjspin}{\hat{\pi}}
\newcommand{\covg}{\overset{\mathfrak{e}}{D}}
\newcommand{\covf}{\overset{\mathfrak{f}}{D}}

\newcommand{\constonee}{\accentset{\mathfrak{e}}{C}_1}
\newcommand{\consttwoe}{\accentset{\mathfrak{e}}{C}_2}
\newcommand{\constthreee}{\accentset{\mathfrak{e}}{C}_3}
\newcommand{\constonef}{\accentset{\mathfrak{f}}{C}_1}
\newcommand{\consttwof}{\accentset{\mathfrak{f}}{C}_2}
\newcommand{\constthreef}{\accentset{\mathfrak{f}}{C}_3}
\newcommand{\constonekin}{\accentset{\textrm{\ae}}{C}_1}
\newcommand{\consttwokin}{\accentset{\textrm{\ae}}{C}_2}
\newcommand{\constthreekin}{\accentset{\textrm{\ae}}{C}_3}

\newcommand{\kinveccoeffimposeconstr}{\accentset{\textrm{\ae}}{A}_\mathcal{V}}
\newcommand{\kinantcoeffimposeconstr}{\accentset{\textrm{\ae}}{A}_\mathcal{A}}
\newcommand{\kinsymcoeffimposeconstr}{\accentset{\textrm{\ae}}{A}_\mathcal{S}}
\newcommand{\kintrcoeffimposeconstr}{\accentset{\textrm{\ae}}{A}_\mathcal{T}}
\newcommand{\kincoeffimposeconstr}{\accentset{\textrm{\ae}}{A}_\mathcal{I}}
\newcommand{\kinveccoeffimposeconstrinv}{\accentset{\textrm{\ae}}{B}_\mathcal{V}}
\newcommand{\kinantcoeffimposeconstrinv}{\accentset{\textrm{\ae}}{B}_\mathcal{A}}
\newcommand{\kinsymcoeffimposeconstrinv}{\accentset{\textrm{\ae}}{B}_\mathcal{S}}
\newcommand{\kintrcoeffimposeconstrinv}{\accentset{\textrm{\ae}}{B}_\mathcal{T}}
\newcommand{\kincoeffimposeconstrinv}{\accentset{\textrm{\ae}}{B}_\mathcal{I}}

\newcommand{\aux}{a}
\newcommand{\tetrade}{\rotatebox[origin=c]{180}{$\mathfrak{e}$}}
\newcommand{\tetradf}{\rotatebox[origin=c]{180}{$\mathfrak{f}$}}
\newcommand{\tetradmean}{\rotatebox[origin=c]{180}{\text{\ae}}}
\newcommand{\cotetradmean}{\text{\ae}}
\newcommand{\sourceg}{\overset{\mathfrak{e}}{S}}
\newcommand{\sourcef}{\overset{\mathfrak{f}}{S}}
\newcommand{\detg}{\mathfrak{e}}
\newcommand{\detf}{\mathfrak{f}}

\newcommand{\Mkin}{\accentset{\textrm{\ae}}{M}}

\begin{document}
	\title{Teleparallel bigravity}

	\author{Daniel Blixt}
	\email{d.blixt@ssmeridionale.it}
	\affiliation{Scuola Superiore Merdionale, Largo S. Marcellino 10, I-80138, Napoli, Italy}

	\author{Manuel Hohmann}
	\email{manuel.hohmann@ut.ee}
	\affiliation{Laboratory of Theoretical Physics, Institute of Physics,  University of Tartu, W. Ostwaldi 1, 50411 Tartu, Estonia}

	\author{Tomi Koivisto}
	\email{tomi.koivisto@ut.ee}
	\affiliation{Laboratory of Theoretical Physics, Institute of Physics, University of Tartu, W. Ostwaldi 1, 50411 Tartu, Estonia}
	\affiliation{National Institute of Chemical Physics and Biophysics, R\"avala pst. 10, 10143 Tallinn, Estonia}

	\author{Luca Marzola}
	\email{luca.marzola@cern.ch}
	\affiliation{National Institute of Chemical Physics and Biophysics, R\"avala pst. 10, 10143 Tallinn, Estonia}

	\begin{abstract}
	We write down the teleparallel equivalent to Hassan-Rosen bigravity, which is written using a torsionful but curvature-free connection. The theories only differ by a boundary term. The equivalence was proven, both by using perturbation theory and Hamiltonian analysis. It is further shown how one can construct novel bigravity theories within the teleparallel framework. Some of those are analyzed through perturbation theory, and it is found that all of the considered novel bigravity theories suffer from pathologies. In particular, it is found that a construction with two copies of new general relativity leads to ghostly degrees of freedom which are not present in the single tetrad teleparallel corresponding theory. We demonstrate how the teleparallel framework allows to easily create theories with derivative interaction. However, it is shown through perturbation theory that the simplest model is not viable. Furthermore, we demonstrate some steps in the Hamiltonian analysis of teleparallel bigravity with two copies of new general relativity and some toy models. The results rule out some of the novel teleparallel bigravity theories, but also demonstrate techniques in perturbation theory and Hamiltonian analysis which could be further used for more profound theories in the future.
	\end{abstract}

	\maketitle

	\section{Introduction}
	Quantum field theory has successfully led to a consistent description of all fundamental forces of nature within the standard model of particle physics, except gravity. The standard model in its present form has passed numerous experimental tests, including most prominently the discovery of the Higgs boson almost half a century after its prediction. Modifying such a robust theory, to make it consistent with general relativity seems very difficult. A more natural approach would be to look at modifications of general relativity. The gauge structure encountered in the standard model and unexplained observations in Nature guide us towards the most motivated ways to explore theories beyond general relativity.
	\\ \indent General relativity has also been very successful in passing experimental tests. It has successfully predicted the deflection of light from the sun and the precession of Mercury. One may ask why we should consider theories beyond general relativity. Beside a philosophically appealing idea, that there would exist a theory for everything, we need a quantum theory of gravity in order to understand the early universe and black holes. However, general relativity is a notoriously non-renormalizable theory, as is very nicely explained in \cite{Zee:2003mt}, so that the usual quantization methods used in the standard model cannot be applied.
	\\ \indent From observations at scales of galaxies and larger, we conclusively find evidence for the so-called dark sector of the universe. The dark sector consists of dark matter and dark energy, which make up about 95\% of our observable universe. Dark energy gives rise to an expanding universe at an accelerating speed. The existence of dark matter can be inferred indirectly by its gravitational effects on the motion of stars within galaxies, gravitational lensing, and structure formation. However, neither of these components has been observed in any laboratory experiment searching for a suitable extension of the standard model. Since gravity is the dominant force at the scales where dark matter and dark energy are observed, it is motivating to look at extensions of general relativity which could give us an understanding of the dark sector of the universe.
	\\ \indent The first question to ask before investigating modifications of general relativity is ``which formulation of general relativity should be modified?''. In \cite{BeltranJimenez:2019tjy} the different formulations of general relativity are explained. Einstein already proposed an alternative formulation of general relativity, which is nowadays called ``teleparallel equivalent to general relativity''\cite{doi:10.1002/3527608958.ch36,Maluf:2013gaa}, or ``TEGR''. However, the third equivalent formulation called ``symmetric teleparallel equivalent to general relativity'', or ``STEGR'', was not found until 1998 in \cite{Nester:1998mp} and only more recently unified into the general teleparallel theory in \cite{BeltranJimenez:2019odq}. In this article we will work in the framework of ``teleparallel gravity'', where our starting point will be the teleparallel equivalent to general relativity. For reviews about teleparallel gravity, see \cite{AP,Bahamonde:2018rgz,Bahamonde:2021zah,Bahamonde:2021gfp,Blixt:2020ekl,Krssak:2018ywd}. For literature about the viability of teleparallel gravity theories, see \cite{Bahamonde:2022ohm,BeltranJimenez:2020fvy,Blagojevic:2000qs,Blagojevic:2020dyq,Blixt:2018znp,Blixt:2019ene,Blixt:2019mkt,Capozziello:2023foy,Chen:2014qtl,Cheng:1988zg,Ferraro:2016wht,Ferraro:2018tpu,Ferraro:2020tqk,Golovnev:2018wbh,Golovnev:2019kcf,Golovnev:2020aon,Golovnev:2017dox,Golovnev:2020nln,Hohmann:2019sys,Hohmann:2019gmt,BeltranJimenez:2019nns,Koivisto:2018loq,Krssak:2015oua,Kuhfuss:1986rb,Li:2011rn,Maluf:2000ag,Maluf:1994ji,Mitric:2019rop,Nester:2017wau,Okolow:2011np,Okolow:2011nq,Okolow:2013lwa,Ong:2013qja,Ortin:2015hya,PhysRevD.19.3524,VanNieuwenhuizen:1973fi}. Teleparallel gravity, in contrast to the conventional curvature-based formulation of gravity in terms of the Einstein-Hilbert Lagrangian, is a gauge theory of gravity with a Yang-Mills-like structure\footnote{The alleged similarity with Yang-Mills is though compromised by the degeneracy of the gauge group, the inconsistency of the minimal coupling, the need for a non-canonical Lagrangian, the violation of Yang-Mills Bianchi identity {\it etc}. There is, however, a new approach to gauging translations which may alleviate these issues \cite{Koivisto:2019ejt,Koivisto:2022uvd}.\label{caveat}} \cite{Blagojevic:2013xpa,Cho:1975dh}, similar to that of the standard model. Thus, teleparallel gravity theories are more reminiscent of the standard model of particle physics.\\
	\indent The second question to ask is ``how to modify gravity?''. In this article we want to investigate the possibility of adding a massive spin-2 field. Already in 1939 Fierz and Pauli addressed the question of a  possible formulation of a massive gravity theory \cite{Pauli:1939xp}. There were some further historical attempts, such as \cite{Boulware:1973my,VanNieuwenhuizen:1973fi}, but not until 2010 a consistent theory of massive gravity was formulated which avoids the previously encountered problems of ghost instabilities \cite{deRham:2010kj}. Shortly after Hassan and Rosen wrote down a consistent gravitational theory for a massless and a massive spin-2 field \cite{Hassan:2011zd}. This theory was named ``bimetric gravity'' but is sometimes referred as ``bigravity'' or ``bimetric theory''. For a review about bimetric gravity, see \cite{Schmidt-May:2015vnx,Heisenberg:2021dqm}. For work on the consistency of bigravity theories, see \cite{Hassan:2013pca,Hinterbichler:2012cn,Li:2015iwc,Tukhashvili:2017eqc,Hassan:2014gta,Hassan:2017ugh,Kocic:2018yvr,Kocic:2019ahm,Kocic:2019axa,Schmidt-May:2016hsx,Hassan:2012rq,Hassan:2012gz,Hassan:2012wr,Hassan:2012qv,Hassan:2018mbl,Molaee:2018brt,Higuchi:1986py,Hassan:2018mcw,Soloviev:2022qlh}. Bimetric gravity has the potential of describing dark matter \cite{Babichev:2016bxi,Manita:2022tkl} and viable cosmological solutions \cite{Akrami:2012vf,Akrami:2013pna,Bassi:2023ymf,Hogas:2019ywm,Hogas:2021saw,Hogas:2022owf,Solomon:2014dua,Luben:2018ekw,Luben:2019yyx}. The aim of this article is to investigate if we can write down viable teleparallel bigravity theories, that is, to formulate a self-consistent theory of two dynamical tetrads using a flat, metric compatible, but torsionful connection.\\
	\indent  A promising feature coming from the structure of teleparallel bigravity is the possibility to construct terms with derivative interactions which are much more symmetric than those constructed in curvature based theories of bigravity (see for example \cite{Li:2015iwc,Li:2015izu,Tukhashvili:2017eqc}). Derivative interaction has gained interest in bigravity since they are related to theories with conformal symmetries \cite{Apolo:2016ort,Apolo:2016vkn}. These theories exhibit a symmetry which manifests with the existence of a partially massless spin-2 field instead of the massive spin-2 field. There is, however, a no-go theorem\footnote{This is rather a conjecture, since the argument only holds up to cubic order, whereas the interacting theory might only start at quartic order, as the authors stress in their conclusions.} forbidding self-interactions for a single partially massless spin-2 field \cite{Garcia-Saenz:2015mqi}. This no-go theorem is here avoided, by the fact that this is a theory of several interacting fields as in \cite{Boulanger:2019zic}.
	\\ \indent We provide a new teleparallel equivalent formulation for Hassan-Rosen bigravity (TEHR) where curvature is replaced by torsion analogously to the teleparallel formulation of general relativity. This provides a theory of a massless and a massive spin-2 field, which, in addition, more evidently exhibits a Yang-Mills-like gauge structure. These properties make the theory more reminiscent of the standard model of particle physics compared to the formulation of general relativity the curvature formulation of the theory of general relativity (see though footnote \ref{caveat} in this paper and the first paragraph in the paper \cite{Koivisto:2019ejt}).
	\\ \indent We show that the Hamiltonian of TEHR has the same structure as in \cite{Hinterbichler:2012cn} so we expect the number of degrees of freedom to be consistent. Furthermore, the linear analysis around a Minkowski background describes massless and massive spin-2 fields just as in Hassan-Rosen bimetric gravity. This formulation will work as a conceptually different view of Hassan-Rosen bimetric gravity, as well as a new base for possible extensions of the theory. Furthermore, we present a very general class of novel theories within the TEHR framework and look at the linear level of some interesting cases and use the Dirac-Bergmann algorithm to express the Hamiltonians for some examples.\\
	\indent In section \ref{sec:TeleparallelBigravity}, we describe the theories of bigravity and teleparallel gravity to finally introduce our novel theory of teleparallel bigravity. In section \ref{sec:Linear} we make the linear analysis for a large subclass of teleparallel bigravity, and in section \ref{sec:Nonlinear} we derive the primary constraints for these theories and the Hamiltonian for two specific examples.\\
	\indent We adopt to the following conventions: metrics follow the $\eta_{AB}=\mathrm{diag}(-1,1,1,1)$ sign convention. Greek indices denote global coordinate indices, capital Latin indices represent Lorentz indices, and small Latin indices denote spatial coordinate indices. We use the notation for symmetrizing indices of a tensor by $2A_{(\mu\nu)}:=A_{\mu\nu}+A_{\nu\mu}$ and, similarly, for antisymmetrizing indices of a tensor we have $2A_{[\mu\nu]}:=A_{\mu\nu}-A_{\nu\mu}$.

	\section{Teleparallel bigravity}
	\label{sec:TeleparallelBigravity}
	In this section we formulate a novel class of theories of gravity which we choose to call teleparallel bigravity. The basic idea of this theory is quite simple. By taking the Hassan-Rosen bimetric gravity theory as our starting point, which is based on a formulation with a metric compatible and torsion-free connection, one can view this action in the teleparallel framework. In this formulation we are able to construct novel theories of gravity, as well as recovering equivalent theories, such as the teleparallel equivalent to Hassan-Rosen bimetric gravity. In order to reach a construction of teleparallel bigravity we start by reviewing Hassan-Rosen bimetric gravity in subsection \ref{sec:HR}. Then, in order to clarify the distinction between the idea of teleparallel bigravity and other novel theories with names which suggest very similar ideas, we devote section \ref{sec:Othertheories} to shortly review these theories. In section \ref{sec:TEGR} we give a short review for teleparallel equivalent to general relativity, and its extension to new general relativity is reviewed in section \ref{sec:NGR}. Section \ref{sec:teleparallelbigravity} is devoted to the introduction of teleparallel bigravity.

	\subsection{Hassan-Rosen bimetric gravity}

	\label{sec:HR}

	Bimetric gravity (or bigravity, bimetric relativity, or bimetric theory) is an extension to general relativity where one metric is added to the theory and is given dynamics. It is a theory of two interacting spin-2 fields. One of the fields is massless (as in general relativity) and the other is massive. In \cite{Hassan:2011zd} the unique ghost-free bimetric gravity theory with the kinetic terms being two copies of the Einstein-Hilbert action is presented. We refer to this theory as Hassan-Rosen bimetric gravity, and its corresponding action is
	\begin{align}
	\label{HRmetric}
	\begin{split}
	S_{\mathrm{HR}}=&m^{2}_{e} \int \mathrm{d}^{4}x \sqrt{-\gee}R\left(\gee\right)+m^{2}_{f}\int \mathrm{d}^{4}x\sqrt{-\gf}R\left(\gf\right)-2m^{4}\int \mathrm{d}^{4}x \sqrt{-\gee}\sum_{n=0}^{4}\beta_{n}e_{n}\left(\sqrt{\gee^{-1}\gf} \right),
	\end{split}
	\end{align}
	where, $\gee$ and $\gf$ are two dynamical metrics (with corresponding tetrads $\ee, \ff$ to be defined later), and $\beta_{n}$ are arbitrary parameters in front of the elementary symmetric polynomials of the eigenvalues  $\lambda_i$: of the matrix $\sqrt{\gee^{-1}\gf}$, given by
	\begin{align}
	e_{0}\left(\sqrt{\gee^{-1}\gf} \right)=1,
	\\ e_{1}\left(\sqrt{\gee^{-1}\gf} \right)=\lambda_{1}+\lambda_{2}+\lambda_{3}+\lambda_{4},
	\\ e_{2}\left(\sqrt{\gee^{-1}\gf} \right)=\lambda_{1}\lambda_{2}+\lambda_{1}\lambda_{3}+\lambda_{1}\lambda_{4}+\lambda_{2}\lambda_{3}+\lambda_{2}\lambda_{4}+\lambda_{3}\lambda_{4},
	\\ e_{3}\left(\sqrt{\gee^{-1}\gf} \right)=\lambda_{1}\lambda_{2}\lambda_{3}+\lambda_{1}\lambda_{2}\lambda_{4}+\lambda_{1}\lambda_{3}\lambda_{4}+\lambda_{2}\lambda_{3}\lambda_{4},
	\\ e_{4}\left(\sqrt{\gee^{-1}\gf} \right)=\lambda_{1}\lambda_{2}\lambda_{3}\lambda_{4}.
	\end{align}
	As in general relativity
	\begin{align}
	\sqrt{-\gee}:=\sqrt{-\mathrm{det}(\gee)},
	\end{align}
	and $R(\gee)$ is the Ricci scalar of the Levi-Civita connection, which essentially consists of up to second order derivatives of the metric $\gee_{\mu\nu}$. The parameters $m, \ m_{\ee}, \ m_{\ff}$ have a dimension of mass and are needed to make the action dimensionless.
	\\ \indent Hassan-Rosen bimetric gravity can also be formulated in the vielbein (or tetrad) formalism \cite{Hinterbichler:2012cn}. In this formalism the action becomes:
	\begin{align}
	\label{HRtetrad}
	\begin{split}
	S_{\mathrm{HR}}=&m^{2}_{\ee}\int \mathrm{d}x^{4}\epsilon_{ABCD}E^A\wedge E^B \wedge R^{CD}(E)+m^{2}_{\ff}\int \mathrm{d}^{4} x \epsilon_{ABCD}F^A\wedge F^B \wedge R^{CD}(F) -
	\\ &-2m^{4}\int  \epsilon_{ABCD}\left[ \beta_{0}E^A\wedge E^B\wedge E^C \wedge E^D+\beta_{1}E^A\wedge E^B\wedge E^C \wedge F^D+ \right.
	\\ &\left. +\beta_{2} E^A\wedge E^B \wedge F^C \wedge F^D+\beta_{3} E^A\wedge F^B  \wedge F^C \wedge F^D+\beta_{4}F^A\wedge F^B \wedge F^C\wedge F^D \right],
	\end{split}
	\end{align}
	where $R^{CD}(E)$ and $R^{CD}(F)$ are the curvature 2-forms of the Levi-Civita connection constructed from the cotetrads $E^A$ and $F^A$ respectively, which relates to the metrics $\gee$ and $\gf$ in the following way:
	\begin{align}
	\label{vierbeing}
	E^A=\ee^A{}_\mu\mathrm{d}x^\mu,\\ \label{vierbeinf}
	F^A=\ff^A{}_\mu\mathrm{d}x^\mu,\\ \label{metricg}
	\gee_{\mu\nu}=\ee^{A}{}_\mu\eta_{AB}\ee^{B}{}_\nu,
	\\ \label{metricf} \gf_{\mu\nu}=\ff^{A}{}_\mu\eta_{AB}\ff^{B}{}_\nu.
	\end{align}
	We also define the tetrads $\tetrade_A{}^\mu$, and $\tetradf_A{}^\mu$, dual to the cotetrads $\ee^A{}_\mu$, and $\ff^A{}_\mu$ respectively. For objects which are not metrics, we may suppress metrics and place indices in the wrong position in order to make expressions look more compact. For instance, if we have a cotetrad $\ee^A{}_\mu$, we may, instead of $\ee^B{}_\nu\eta_{AB}\gee^{\mu\nu}$, write $\ee_A{}^\mu$. This takes the correct index positions as the tetrad, but it should not be confused with the tetrad which we, in this case, denote by $\tetrade_{A}{}^\mu$. Note that the action is written slightly differently from \cite{Hinterbichler:2012cn} and rather as in \cite{Li:2015iwc}. This will then only be equivalent to (\ref{HRmetric}) up to a reparametrisation of the coefficients $\beta_{n}$.

	\subsection{Other bigravity theories}
	\label{sec:Othertheories}
	Our aim is to look at bigravity theories in the teleparallel framework, which assumes vanishing curvature. This includes novel theories of gravity. Recently, other novel bigravity theories have been proposed in articles where either the title is reminiscent to the concept of teleparallel gravity, or the structure of the theory is similar. Before going into the theory of teleparallel gravity, we highlight the main differences between these bigravity theories and teleparallel bigravity, in order to avoid any confusion. In \cite{Damour:2019oru,Nikiforova:2020fbz,Nikiforova:2022pqy} a novel theory named torsion bigravity has been considered. Here, a theory of both curvature and torsion is formulated, where a massless spin-2 field is mediated by the curvature, whereas the massive spin-2 field is mediated by the torsion, which avoids the notion of two metrics (or tetrads) in the construction of a bigravity theory. \\
	\indent In \cite{Markou:2018mhq} a bigravity theory is constructed with vierbeins as the fundamental field in order to yield an antisymmetric part of the field equations. It was motivated by the fact that antisymmetric field equations played a big role in the quantization of string theory, which suggests that it would make sense to include antisymmetric parts in a gravity theory. Their construction differs from ours by being curvature-based. However, we cannot exclude cases where the field equations might be equivalent. This theory was extended from the single vierbein massive gravity theory proposed in \cite{Markou:2018wxo}. In \cite{Alexandrov:2019dxm} the structure of the action seems to yield a physics similar to that of our theory, and even closer to that of \cite{Li:2015iwc}. The crucial difference with respect to our approach is that our construction uses more than one connection. In teleparallel bigravity the action contains two connections (corresponding to the two tetrad fields), but only one spin connection. The construction of a bigravity theory with only one connection was found to yield one graviton, instead of two as in other bigravity theories, making this scheme significantly different from teleparllel bigravity\cite{Alexandrov:2019dxm}.

	\subsection{Teleparallel equivalent to general relativity}
	\label{sec:TEGR}
	General relativity is formulated with the Ricci scalar which depends on the spin connection, which is taken to be torsion-free. However, an equivalent formulation can be made by setting the spin connection to be curvature-free and replacing curvature with torsion. The resulting action reads:
	\begin{equation}\label{TPgravity}
	S_{\mathrm{TEGR}}=\int \mathrm{d}^4 x  \mathcal{L}_{\mathrm{TEGR}}=-m_e^2\int \mathrm{d}^4 x \detg \accentset{\ee}{T}^{\rho\mu\nu}\left(\frac{1}{4}\accentset{\ee}{T}_{\rho\mu\nu}+\frac{1}{2}\accentset{\ee}{T}_{\nu\mu \rho}-\gee_{\rho\nu}\accentset{\ee}{T}_\mu\right),
	\end{equation}
	where  $\detg=\det(\ee^A{}_\mu)$, $\accentset{\ee}{T}_\mu=\accentset{\ee}{T}^\nu{}_{\mu\nu}$,  $\accentset{\ee}{T}^\alpha{}_{\mu\nu}=\tetrade_A{}^\alpha \accentset{\ee}{T}^A{}_{\mu\nu}$ and $\accentset{\ee}{T}^A{}_{\mu\nu}$ are the coefficients of the two-form
	\begin{align} \label{eq:torsionOneformE}
	\mathbb{\accentset{\ee}{T}}^A=\frac{1}{2}\accentset{\ee}{T}^A{}_{\mu\nu} \mathrm{d}x^\mu \wedge \mathrm{d}x^\nu ; \quad \accentset{\ee}{T}^A{}_{\mu\nu}=2 \partial_{[\mu}\ee^A{}_{\nu]} =\partial_{\mu}\ee^A{}_\nu-\partial_{\nu}\ee^A{}_\mu,
	\end{align}
	and where, for our purposes, we chose to set a priori the so-called spin connection to zero. In particular, it is known that this works for the teleparallel equivalent of general relativity and its most popular modifications \cite{Blixt:2022rpl,Blixt:2018znp,Golovnev:2021omn,Golovnev:2023yla}.\\
	\indent This action is equivalent to the Einstein-Hilbert action modulo a boundary term \cite{AP,Maluf:2013gaa} (which does not appear in this theory). This can be seen by realizing that the Ricci scalar of the teleparallel connection which vanishes due to the flatness condition can be written as
	\begin{align}
	0\equiv\accentset{\ee}{\mathcal{R}}=\accentset{\ee}{R}+\accentset{\ee}{T}-\frac{2}{\detg}\partial_{\mu}\left(\detg\accentset{\ee}{T}^\mu\right),
	\end{align}
	where $\accentset{\ee}{R}$ denotes the curvature of the torsion-free Levi-Civita connection and $\LG_{\mathrm{TEGR}} = -m_g^2\accentset{\ee}{T}$. One realize immediately from this relation that
	\begin{align}
	\label{eq:Equivalence}
	S_{\mathrm{EH}}:= m_e^2\int \mathrm{d}^4x \detg \accentset{\ee}{R}=\int \mathrm{d}^4 x \LG_{\mathrm{TEGR}} +2m_e^2\int \mathrm{d}^4 x \partial_\mu \left(\detg\accentset{\ee}{T}^\mu\right)=S_\mathrm{TEGR}+B.T.,
	\end{align}
	where $B.T.$ denotes the apparent boundary term. From this equation we identify the curvature of the Levi-Civita connection and realize that in teleparallel gravity it is equivalent to \eqref{TPgravity} up to a boundary term.

	\subsection{New general relativity}
	\label{sec:NGR}
	Considering \eqref{TPgravity} it is apparent that one easily can formulate more theories quadratic in the torsion than the special case equivalent to general relativity. By relaxing the coefficients in \eqref{TPgravity} to be arbitrary constants, we get a generalized theory of gravity. It appears that any other term quadratic in the torsion tensor and without any derivatives acting on the torsion would violate parity \cite{PhysRevD.19.3524}. Our starting point is then
	\begin{equation}
	\label{eq:NGRAction}
	S_{\mathrm{NGR}}=\int \mathrm{d}^4 x  \mathcal{L}_{\mathrm{NGR}}=-m_e^2\int \mathrm{d}^4 x \detg \accentset{\ee}{T}^{\rho\mu\nu}\left(\constonee \accentset{\ee}{T}_{\rho\mu\nu}+\consttwoe \accentset{\ee}{T}_{\nu\mu \rho}+\constthreee g_{\rho\nu}\accentset{\ee}{T}_\mu\right).
	\end{equation}
	This class of theories have extensively been studied in \cite{Blixt:2018znp,Blagojevic:2000qs,Okolow:2011np,PhysRevD.19.3524,Koivisto:2018loq,Blixt:2019ene,BeltranJimenez:2019nns,Hohmann:2019sys,Hohmann:2018jso,Ualikhanova:2019ygl,Cheng:1988zg}. Sometimes this theory is named new general relativity, however, the original name was reserved for only a restricted class by forcing $2\constonee+\consttwoe+\constthreee=0$ \cite{PhysRevD.19.3524}. It is known that if one does not force this restriction the parameterized post Newtonian parameters will deviate from those of general relativity \cite{PhysRevD.19.3524,Ualikhanova:2019ygl} and ghost instabilities are bound to arise \cite{BeltranJimenez:2019nns,Koivisto:2018loq,Kuhfuss:1986rb,Ortin:2015hya}. The ghost instabilities will not get removed in the extension to teleparallel bigravity. Since our goal is to make some prior investigation for what novel teleparallel bigravity theories can be viable, this restriction makes sense. However, to canonically write the action for this restricted case one should add a Lagrange multiplier corresponding to the primary constraints imposed by this restriction to \eqref{eq:NGRAction}\footnote{And as well one should not put the spin connection to vanish beforehand, however, this is shown in appendix \ref{appendix:GaugeFixing} to not affect the counting of the number of degrees of freedom.}. Despite this naming, it makes sense to not fix the coefficients $\constonee, \ \consttwoe$, and $\constthreee$ beforehand since we can then realize how the choice of parameters has to be chosen to remove ghost-instabilities.

	\subsection{Teleparallel bigravity}
	\label{sec:teleparallelbigravity}
	Hassan-Rosen bimetric gravity is given by the action of either (\ref{HRmetric}) or (\ref{HRtetrad}). It has two Einstein-Hilbert terms, one for each metric, and an interaction potential. Many equations are similar to general relativity, with the addition of an analogous equation depending on the second metric.
	\\ \indent In this section we will consider a new framework of gravity theories which we call ``teleparallel bigravity''. We write down a general action quadratic in the torsion and that does not violate parity. Then we count degrees of freedom for the special case of ``teleparallel equivalent to Hassan-Rosen bigravity'', showing that the theory is a ghost-free theory of massless and massive spin two fields.

	\subsubsection{Novel terms with derivative interactions}
	\label{sec:Novelterms}
	Bigravity theories with derivative interactions have been considered in \cite{Apolo:2016ort,Tukhashvili:2017eqc,Li:2015iwc}. The naïve actions provided in \cite{Li:2015iwc,Li:2015izu,deRham:2015rxa} generically have ghost instabilities due to a relative sign and the full Hamiltonian analysis of \cite{Apolo:2016ort} have not been carried out. The teleparallel gravity framework, however, gives much more freedom to create bigravity theories with derivative interactions. In this section we explore if examples that are potentially ghost-free can be created. The most general action which is quadratic in torsion in this framework reads 
	\begin{align}
	\label{GenBiGrav}
	S_{\mathrm{TBG}}=-\int \mathrm{d}^4x\left[m_e^2\accentset{\ee}{T}^A{}_{\mu\nu}\accentset{\ee}{T}^B{}_{\rho \sigma}\overset{\ee}{G}_{AB}{}^{\mu\nu\rho\sigma}+m_{ef}^2\accentset{\ee}{T}^A{}_{\mu\nu}\accentset{\ff}{T}^B{}_{\rho \sigma}\overset{kin}{G}{}_{AB}{}^{\mu\nu\rho\sigma}+m_f^2\accentset{\ff}{T}^A{}_{\mu\nu}\accentset{\ff}{T}^B{}_{\rho \sigma}\overset{\ff}{G}_{AB}{}^{\mu\nu\rho\sigma}\right]+V(E,F),
	\end{align}
	where $V(E,F)$ is an interaction potential, and $\overset{\ee}{G}_{AB}{}^{\mu\nu\rho\sigma}, \ \overset{kin}{G}_{AB}{}^{\mu\nu\rho\sigma},$ and $\overset{\ff}{G}_{AB}{}^{\mu\nu\rho\sigma}$ are some combinations of tetrads and metrics not containing any derivatives. Analogously to new general relativity, we introduced
	\begin{align} \label{eq:TorsionOneformF}
	\mathbb{\accentset{\ff}{T}}^A=\frac{1}{2}\accentset{\ff}{T}^A{}_{\mu\nu} \mathrm{d}x^\mu \wedge \mathrm{d}x^\nu ; \quad \accentset{\ff}{T}^A{}_{\mu\nu}=2 \partial_{[\mu}\ff^A{}_{\nu]} =\partial_{\mu}\ff^A{}_\nu-\partial_{\nu}\ff^A{}_\mu,
	\end{align}
	where we again set the spin connection to zero, which we have shown to not affect the number of degrees of freedom in appendix \ref{appendix:GaugeFixing}.
	Below, we display a figure for teleperallel bigravity and its limits.

	\tikzstyle{decision} = [diamond, draw, fill=blue!20,
	text width=4.5em, text badly centered, node distance=2cm, inner sep=0pt]
	\tikzstyle{block} = [rectangle, draw, fill=green!40,
	text width=7em, text centered, rounded corners, minimum height=3em]
	\tikzstyle{ma} = [ellipse, draw, fill=purple!12,
	text width=6em, text centered, rounded corners, minimum height=2.5em]
	\tikzstyle{bg} = [ellipse, draw, fill=red!32,
	text width=6em, text centered, rounded corners, minimum height=2.5em]
	\tikzstyle{tbg} = [ellipse, draw, fill=yellow!32,
	text width=6em, text centered, rounded corners, minimum height=2.5em]
	\tikzstyle{novel} = [ellipse, draw, fill=purple!12,
	text width=6em, text centered, rounded corners, minimum height=2.5em]
	\tikzstyle{ngr} = [ellipse, draw, fill=blue!12,
	text width=6em, text centered, rounded corners, minimum height=2.5em]
	\tikzstyle{mink} = [ellipse, draw, fill=blue!32,
	text width=6em, text centered, rounded corners, minimum height=2.5em, node distance=15cm]
	\tikzstyle{line} = [draw, -latex']
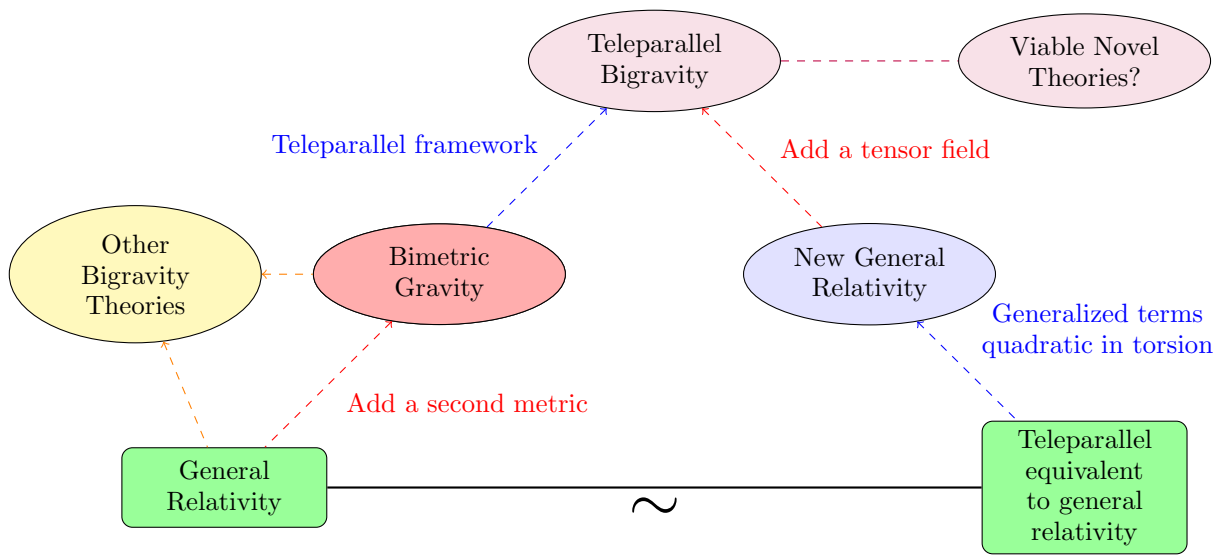
	\begin{center}
		\begin{tikzpicture}[node distance = 4.0cm, auto]
		\node [ma] (TB) {Teleparallel Bigravity };
		\node [ngr, below right of=TB ] (NGR) {New General Relativity};
		\node [bg, below left of=TB ] (BG) {Bimetric Gravity};
		\node [tbg, left of=BG ] (TBG) {Other Bigravity Theories};
		\node [bg, below left of=TB ] (BG) {Bimetric Gravity};
		\node [novel, above right of=NGR ] (NOVEL) {Viable Novel Theories?};
		\node [block, below right of=NGR ] (TG) {Teleparallel equivalent to general relativity};
		\node [block, below left of=BG ] (GR) {General Relativity};
		\draw[<-, dashed, red] (TB) -- node  {{ Add a tensor field}} (NGR);
		\draw[->, dashed, blue] (BG) -- node  {{ Teleparallel framework}} (TB);
		\draw[<-, dashed, blue] (NGR) -- node
		{\begin{tabular}{c}
			Generalized terms  \\
			quadratic in torsion \\
			\end{tabular}} (TG);
		\draw[-, dashed, purple] (TB) -- node  {{ }} (NOVEL);
		\draw[<-, dashed, red] (BG) -- node  {{ Add a second metric}} (GR);
		\draw[<-, dashed, orange] (TBG) -- node  {{}} (BG);
		\draw[<-, dashed, orange] (TBG) -- node  {{}} (GR);
		\draw[-, thick] (TG) -- node  {{\Huge $\mathbf{\sim}$}} (GR);
		\end{tikzpicture}
		\vspace{3.2cm}
	\end{center}
	The figure should be understood in the following way. In the lower part in green color we display two equivalent formulations of general relativity. That is, the standard formulation with a metric compatible and torsion free connection and the teleparallel equivalent to general relativity. With the standard formulation of general relativity, it is known from the literature that one can add a second metric and make it dynamical to create a bimetric gravity theory (following the red arrow in the figure). We note that the concept bigravity can be used for more theories than just the standard Hassan-Rosen bimetric gravity, with some of them briefly discussed in section \ref{sec:Othertheories}. From bimetric gravity, one can first look into the tetrad formulation of the theory, and then take inspiration from the known teleparallel gravity theories to construct a teleparallel bigravity theory by changing to the teleparallel framework (blue arrow). \\
	\indent Another logical path for constructing teleparallel bigravity is to start with the teleparallel equivalent to general relativity. By just looking at this action, one can realize that it can be generalized to new general relativity. With new general relativity we can then add a second tetrad which we make dynamical in a very similar way to what was done in the construction of bimetric gravity. This as well leads to the formulation of teleparallel bigravity. Teleparallel bigravity, hence, have the limits of Hassan-Rosen bimetric gravity, new general relativity, and general relativity. In teleparallel bigravity there are novel theories of gravity which may, or may not be viable.

	\subsubsection{Teleparallel equivalent to Hassan-Rosen bigravity}
	\label{sec:EquivalentBigravity}

	We will work in the tetrad formalism since this has been useful in both teleparallel and bimetric gravity.
	We make an ansatz that we simply replace the Einstein-Hilbert terms in (\ref{HRtetrad}) with two copies of (\ref{TPgravity}). We will in section \ref{sec:Linear} show that it reproduces the linear equations of Hassan-Rosen bimetric gravity, and in section \ref{sec:dof} that the counting gives the expected degrees of freedom. The action then is:
	\begin{align}
	\label{TeleBi}
	\begin{split}
	S_{\mathrm{TEHR}}&=-m^{2}_{e}\int \mathrm{d}^{4}x\detg\accentset{\ee}{T}-m^{2}_{f}\int \mathrm{d}^{4} x \detf\accentset{\ff}{T} -
	\\ &-2m^{4}\int \epsilon_{ABCD}\left[ \beta_{0}E^A\wedge E^B\wedge E^C \wedge E^D+\beta_{1}E^A\wedge E^B\wedge E^C \wedge F^D+ \right.
	\\ &\left. +\beta_{2} E^A\wedge E^B \wedge F^C \wedge F^D+\beta_{3} E^A\wedge F^B  \wedge F^C \wedge F^D+\beta_{4}F^A\wedge F^B \wedge F^C\wedge F^D \right].
	\end{split}
	\end{align}
	Here $\detf=\det(\ff^A{}_\mu)$, and, $E^A$, $F^B$, $\gee$, and $\gf$ are defined in \eqref{vierbeing}, \eqref{vierbeinf}, \eqref{metricg}, \eqref{metricf} respectively, $\accentset{\ee}{T}$ is defined in equation (\ref{TPgravity}) and defines $\LG_{\mathrm{TEGR}}$ together with the volume element $\detg$. This is formulated in a gauge where the spin connection vanishes (which is not the canonical formulation). We show in appendix \ref{appendix:GaugeFixing} that the counting of degrees of freedom at the nonlinear level (using Hamiltonian analysis) is independent of our gauge choice. Since this greatly simplifies the Hamiltonian analysis, we choose to set the spin connection to zero. \\
	\indent It is well known that the action given by (\ref{TPgravity}) is equivalent to the Einstein-Hilbert action up to a surface term given by \eqref{eq:Equivalence}.
	Hence, the kinetic part of (\ref{TeleBi}) is equivalent to the kinetic part of Hassan-Rosen bimetric gravity up to a couple of surface terms.
	\\ \indent To get a true equivalence we need to make sure that the theory contains the same number of degrees of freedom as in Hassan-Rosen bimetric gravity. We should also investigate whether the consistency-tests performed in Hassan-Rosen bimetric gravity still holds in this new formulation of the theory.

	\subsubsection{Counting degrees of freedom}
	\label{sec:dof}

	In \cite{Maluf:1994ji,Blagojevic:2000qs,Nester:2017wau} it is shown that the degrees of freedom count for (\ref{TPgravity}) coincides with that of general relativity. Hence the teleparallel formulation of general relativity is a ghost free theory for a spin-2 field as in general relativity.
	\\ \indent We now describe how degrees of freedom are counted in Hassan-Rosen bimetric gravity, since we are looking at an extension to this theory. Hassan-Rosen bimetric gravity is formulated in terms of two dynamical metrics. Note that first class constraints remove 2 degrees of freedom each in phase-space, whereas second class constraints remove only 1. Starting with 20 degrees of freedom for each metric (including conjugate momenta). 8 degrees of freedom can be removed from first class constraints associated with a common diffeomorphism invariance for the two metrics. 4 additional degrees of freedom can be removed from gauge fixing one set of lapse and shift. 4 more are removed by Hamiltonian and momentum constraints, normally realized from the field equations of analogous to \eqref{eq:LTEHR}. One set of lapse and shift appears as Lagrange multipliers which gives us 4 first class primary constraints which removes 8 additional degrees of freedom. Finally, 2 more degrees of freedom are removed by an additional second class constraint with an associated secondary second class constraint. This leaves us with $40-26=14=(2+5)\times 2$ degrees of freedom corresponding to a massless and massive spin-2 field and its corresponding conjugate momenta. Note that the 2 last second class constraints follow from the specific form of the interaction potential \cite{Hassan:2011zd,Hassan:2018mbl} and removes the pathological Boulware-Deser ghost\footnote{Note that in the absence of the interaction potential, there are no connection between the kinetic terms and there will be two sets of diffeomorphism invariances and this will bring us down to two massless spin-2 fields.}. We should, hence, carefully check whether this constraint still remains in our new theory.
	\\ \indent In order to make the Hamiltonian analysis, we need the conjugate momenta, which comes from the fields which have time-derivatives. Hence, we need to distinguish between space and time and make a so-called 3+1 decomposition.\footnote{Note that a valid simultaneous 3+1 decomposition of the two metrics must be done such that they both share a common spacelike hypersurface \cite{Kocic:2018ddp,Kocic:2018yvr,Kocic:2019ahm}.} To avoid cumbersome repetition only the objects associated with one metric ($\gee$) will be displayed, and the analogous parts follows. In the case of TEGR there are 6 additional primary constraints compared the metric GR and they have been shown to be of first class, thus removing the dynamics of all the extra components introduced when treating tetrads as canonical fields instead of the metric. These are only slightly different from the usual tetrad formulation of bigravity \cite{Hinterbichler:2012cn} due to the presence of torsion terms. 
	
		By making a 3+1 decomposition of spacetime in ADM variables we can express the Hamiltonian for teleparallel equivalent to general relativity \cite{Maluf:1994ji,Blixt:2018znp,Blagojevic:2000qs,Nester:2017wau}. Following the notation of \cite{Blixt:2018znp}, but adding an overset label $\ee$ which will be important when extending the result to the teleparallel equivalent to Hassan-Rosen bigravity. The ADM-variables used are $\lapsee$ (lapse), $\shifte^{i}$ (shift), spatial tetrad $\ee^A{}_i$ and its conjugate momenta $\conje_A{}^i$, induced metric $\inducee_{ij}$ and its Levi-Civita covariant derivative $D_i$ and spatial volume element $\sqrt{\inducee}$, a normal vector to constant hypersurfaces $\normee^A=-\frac{1}{6}\epsilon^A{}_{BCD}\ee^B{}_i\ee^C{}_j\ee^D{}_k\epsilon^{ijk}$, and the resulting Hamiltonian reads 
	\begin{align}
	\begin{split}
	\label{TEGR}
	\HG&=\lapsee\sqrt{\inducee}\Bigg[  \frac{1}{8m^2_e}\frac{\conje_A{}^{i}}{\sqrt{\inducee}}\frac{\conje_B{}^{l}}{\sqrt{\inducee}}\inducee^{jk}\inducee_{i(l}\ee^A{}_{k)}\ee^B{}_j-\frac{1}{16m^2_e}\frac{\conje_A{}^i}{\sqrt{\inducee}}\frac{\conje_B{}^j}{\sqrt{\inducee}}\ee^A{}_i\ee^B{}_j-\normee^{A}D_{i}\frac{\conje_A{}^i}{\sqrt{\inducee}}-m^2_e\cdot\spatialee \Bigg]
	\\&+\shifte^{k}\sqrt{\inducee}\Big[ \frac{\conje_B{}^j}{\sqrt{\inducee}}\accentset{\ee}{T}^B{}_{jk} -\ee^A{}_k \covg_{i}\frac{\conje_A{}^i}{\sqrt{\inducee}}\Big]+ \sqrt{\inducee}\covg_{i}\left[\left(\lapsee \normee^{A}+\shifte^{j}\ee^A{}_j \right)  \frac{\conje_A{}^i}{\sqrt{\inducee}}\right]+\lambda^{ij}\accentset{\ee}{\mathcal{P}}_{ij}
	\\& \equiv \lapsee \CHRg+\shifte^i\CHRg_i+\lambda^{ij}\accentset{\ee}{\mathcal{P}}_{ij},
	\end{split}
	\end{align}
	where $\equiv$ signifies that a boundary term have been dropped as discussed in \cite{Pati:2022nwi} and 
		\begin{align}
	\accentset{\ee}{\mathcal{P}}_{ij}:=\begin{bmatrix}
	\CHRVecg_1 & \CHRAng_{12} & \CHRAng_{13}
	\\ \CHRAng_{21} & \CHRVecg_2 & \CHRAng_{23}
	\\ \CHRAng_{31} & \CHRAng_{32} & \CHRVecg_3
	\end{bmatrix},
	\end{align}
	with the 6 primary first class constraints
		\begin{align}
	\CHRVecg^i=-\frac{\conje_A{}^i}{\sqrt{\inducee}}\normee^{A}-2m^2_e\accentset{\ee}{T}^B{}_{jk}\inducee^{ij}\ee_B{}^k \approx 0,
	\end{align}
	and
	\begin{align}
	\CHRAng_{mp}=\frac{\conje_A{}^i}{\sqrt{\inducee}}\inducee_{i[p}\ee^A{}_{m]}+m^2_e\accentset{\ee}{T}^B{}_{pm}\normee_{B} \approx 0,
	\end{align}
	where $\mathcal{V}$ stands for ``vector part'' and $\mathcal{A}$ for ``antisymmetric'' part of the irreducible decomposition under the rotation group as in \cite{Blixt:2018znp,Blagojevic:2000qs}. Note that \cite{Hinterbichler:2012cn} did not find the important $	\CHRVecg^i$ constraints and resorted instead to the field equations to demonstrate Lorentz invariance. The constraint algebra satisfy the Lorentz algebra \cite{Blagojevic:2000qs,Ferraro:2016wht,Maluf:2013gaa} which is essential for realizing they are of first class.
	
	From \eqref{TeleBi} it is straightforward to write down the Lagrangian as the back-transformation of the primary Hamiltonian, for teleparallel equivalent to Hassan-Rosen bigravity in a form like the one used in \cite{Hinterbichler:2012cn}:
	\begin{align}
	\label{eq:LTEHR}
	L_{TEHR}=\int \mathrm{d}x^3 \left(\conje \dot{\ee}+\conjf\dot{\ff}-\lapsee \left(\CHRg+\CHRmg\right)-\lapsef\left(\CHRf+\CHRmf\right)-\shifte^i\left(\CHRg_i+\CHRmg_i\right)-\shiftf^i\left(\CHRf_i+\CHRmf_i\right)-\lambda^{ij}\accentset{\ee}{\mathcal{P}}_{ij}-\tilde{\lambda}^{ij}\accentset{\ff}{\mathcal{P}}_{ij}\right),
	\end{align}
	where $\CHRmg,\CHRmf,\CHRmg_i,\CHRmf_i$  comes, and are defined, from the interaction potential which is linear in lapse and shifts due to the fact that they only appear (and are linear) in $\ee^A{}_0$ and $\ff^A{}_0$ terms and the antisymmetric property of the wedge product ensures that they appear exactly one time in each term as discussed in \cite{Hinterbichler:2012cn}. Now we have reproduced the tetrad formulation of bigravity and we, hence, do not expect any difference in the results of Hassan-Rosen bigravity. In section \ref{sec:EquivalentBigravity} we find that at the linear level we recover the linearized modes of Hassan-Rosen bigravity as expected.

	However, one should be cautious since in the special case of choosing the kinetic action to be two copies of TEGR (and similarly by taking two copies of the tetrad formulation of Einstein-Hilbert terms) the antisymmetric components of the tetrads become non-dynamical \cite{Hassan:2018mcw}. \footnote{This is manifest in section \ref{sec:Linear} where the antisymmetric modes disappear for the TEGR coefficients.} Variation with respect to the antisymmetric components generate the following field equations \cite{Hinterbichler:2012cn,Hassan:2018mcw} which turns out to be constraints. Thus, the antisymmetric components are subject to the following 12 constraints:
	\begin{align}
	\frac{\delta V(E,F)}{\delta\ee^A{}_\mu}\eta^{AB}\tetrade_B{}^\nu-\frac{\delta V(E,F)}{\delta \ee^A{}_\nu}\eta^{AB}\tetrade_B{}^\mu=0,\\
	\frac{\delta V(E,F)}{\delta\ff^A{}_\mu}\eta^{AB}\tetradf_B{}^\nu-\frac{\delta V(E,F)}{\delta \ff^A{}_\nu}\eta^{AB}\tetradf_B{}^\mu=0.
	\end{align}
	Note that this is special for TEHR (and for the tetrad HR formulation of HR bigravity and multigravity extensions \cite{Hassan:2018mcw}) and in general teleparallel bigravity theories the antisymmetric part of the tetrads become dynamical, and the above constraints will be spoiled in the absence of $\accentset{\ee}{\mathcal{P}}_{ij},\accentset{\ff}{\mathcal{P}}_{ij}$, which constrains the antisymmetric kinetic part. Absence of primary constraints for the the antisymmetric sector will reduce the number of constraints, also in the case where at least one of $	\CHRVecg^i,	\CHRAng_{mp}$, or their corresponding $\ff$ constraints, is absent. \\
	\indent We will now turn to the linear analysis of general quadratic teleparallel bigravity theories to confirm the results of Hassan-Rosen bigravity and new general relativity, and further explore what novel theories might be viable. In section \ref{sec:Nonlinear} we will do some preliminary investigation of these novel theories.

	\subsubsection{Explicit examples worth considering}
	\label{sec:Fieldequations}
	In this section we argue for what kind of choice of supermetrics provided in teleparallel bigravity \eqref{GenBiGrav} are worth taking a deeper look into. We assume that the interaction potential $V(E,F)$ is as in Hassan Rosen bigravity. That is,
	\begin{align}
	\begin{split}
	\label{HRpot}
	V(E,F)=-2m^{4}\int \epsilon_{ABCD}\left[ \beta_{0}E^A\wedge E^B\wedge E^C \wedge E^D+\beta_{1}E^A\wedge E^B\wedge E^C \wedge F^D+ \right.
	\\ \left. +\beta_{2} E^A\wedge E^B \wedge F^C \wedge F^D+\beta_{3} E^A\wedge F^B  \wedge F^C \wedge F^D+\beta_{4}F^A\wedge F^B \wedge F^C\wedge F^D \right].
	\end{split}
	\end{align}
	This interaction potential has a clear interpretation when expressed in tetrads. Namely, the terms are the volume forms one can create by general combinations of two tetrads in four dimensions. Furthermore, we have a special case $\accentset{\ee}{G}_{AB}{}^{\mu\nu\rho\sigma}=\accentset{\ee}{G}_{AB}{}^{\mu\nu\rho\sigma}(\ee)$ where the first kinetic term is that of new general relativity. We get another copy of new general relativity by assuming $\accentset{\ff}{G}_{AB}{}^{\mu\nu\rho\sigma}=\accentset{\ff}{G}_{AB}{}^{\mu\nu\rho\sigma}(\ff)$. This is already interesting since these include novel theories for which we already have some insights from the literature of new general relativity and Hassan-Rosen bigravity. For the rest of the article we assume that the interaction potential takes the form as in Hassan-Rosen bigravity, that $\accentset{\ee}{G}_{AB}{}^{\mu\nu\rho\sigma}=\accentset{\ee}{G}_{AB}{}^{\mu\nu\rho\sigma}(\ee)$, and $\accentset{\ff}{G}_{AB}{}^{\mu\nu\rho\sigma}=\accentset{\ff}{G}_{AB}{}^{\mu\nu\rho\sigma}(\ff)$. By including a non-zero $\accentset{kin}{G}_{AB}{}^{\mu\nu\rho\sigma}$ we will get novel terms which can be motivated from its conformal symmetry. This term alone does, however, suffer from ghost-instabilities. However, if one later would consider a similar theory with complex tetrads similar to the constructions in \cite{Apolo:2016ort,Apolo:2016vkn} the theory might turn out to be viable. For the linearized level the explicit expressions of the non-vanishing supermetrics does not play a role in the perturbations we define around a Minkowski background. For the non-linear level we will assume that $\accentset{kin}{G}_{AB}{}^{\mu\nu\rho\sigma}=\accentset{kin}{G}_{AB}{}^{\mu\nu\rho\sigma}(\cotetradmean)$, where we define the cotetrad $\cotetradmean^A{}_\mu$ to satisfy
	\begin{align}
	\cotetradmean^A{}_\mu \eta_{AB}\cotetradmean^B{}_\nu=\gh_{\mu\nu},
	\end{align}
	with its corresponding tetrad denoted by $\tetradmean_A{}^\mu$, and $\gh_{\mu\nu}$ being some new Lorentzian metric.

	 For the Hamiltonian analysis we derive the primary constraints for the theory corresponding to this action. However, we restrict further analysis to the special case $\accentset{\ee}{G}_{AB}{}^{\mu\nu\rho\sigma}=\accentset{\ff}{G}_{AB}{}^{\mu\nu\rho\sigma}=0$, which should be viewed as a toy-model since it contains ghosts. The cotetrad $\cotetradmean^A{}_\mu$  for a bigravity theory should depend on the two other cotetrads $\ee^A{}_\mu$ and $\ff^A{}_\mu$, since the theory describes only two dynamical tensor fields. However, the Hamiltonian analysis can be made without much complications to the point where velocities have been Legendre transformed to conjugate momenta. For this reason we leave $\cotetradmean^A{}_{\mu}$ undetermined until this stage of the analysis. We could already make some speculations of what $\cotetradmean^A{}_\mu$ could be. In order to make the term with derivative interactions completely symmetric in $\ee$ and $\ff$ there are in particular two choices. First one is to assume $\accentset{kin}{G }_{AB}{}^{\mu\nu\rho\sigma}=\accentset{kin}{G}_{AB}{}^{\mu\nu\rho\sigma}(\ee)+\accentset{kin}{G}_{AB}{}^{\mu\nu\rho\sigma}(\ff)$. The second choice is to take it to depend on the so-called geometric mean defined via the relation
	\begin{align}
	\label{eq:geometricmean}
	\gh_{\mu\nu}=\ee^A{}_{(\mu} \ff^B{}_{\nu)}\eta_{AB}=\cotetradmean^A{}_\mu \cotetradmean^B{}_\nu \eta_{AB},
	\end{align}
	where $\gh_{\mu\nu}$ is an intermediate metric between $\gee_{\mu\nu}$ and $\gf_{\mu\nu}$ \cite{Kocic:2018ddp}.
	The explicit expressions for the supermetrics are,
	\begin{align}
	\label{eq:supmetee}	\begin{split}
	\accentset{\ee}{G}_{AB}{}^{\mu\nu\rho\sigma}(\ee)&=\constonee\eta_{AB}\eta^{CD}\eta^{EF}\tetrade_C{}^\rho \tetrade_D{}^{[\mu}\tetrade_F{}^{\nu]}\tetrade_F{}^{\sigma}-\consttwoe\eta^{CD}\tetrade_B{}^{[\mu}\tetrade_C{}^{\nu]}\tetrade_D{}^{[\rho}\tetrade_A{}^{\sigma]}-\constthreee\eta^{CD}\tetrade_A{}^{[\mu}\tetrade_C{}^{\nu]}\tetrade_D{}^{[\rho}\tetrade_B{}^{\sigma]}
	\end{split},\\
	\label{eq:supmetkin}	\begin{split}
	\accentset{kin}{G}_{AB}{}^{\mu\nu\rho\sigma}(\cotetradmean)&=\constonekin\eta_{AB}\eta^{CD}\eta^{EF}\tetradmean_C{}^\rho \tetradmean_D{}^{[\mu}\tetradmean_F{}^{\nu]}\tetradmean_F{}^{\sigma}-\consttwokin\eta^{CD}\tetradmean_B{}^{[\mu}\tetradmean_C{}^{\nu]}\tetradmean_D{}^{[\rho}\tetradmean_A{}^{\sigma]}\\
	&-\constthreekin\eta^{CD}\tetradmean_A{}^{[\mu}\tetradmean_C{}^{\nu]}\tetradmean_D{}^{[\rho}\tetradmean_B{}^{\sigma]}
	\end{split},\\
	\label{eq:supmetff}	\begin{split}
	\accentset{\ff}{G}_{AB}{}^{\mu\nu\rho\sigma}(\ff)&=\constonef\eta_{AB}\eta^{CD}\eta^{EF}\tetradf_C{}^\rho \tetradf_D{}^{[\mu}\tetradf_F{}^{\nu]}\tetradf_F{}^{\sigma}-\consttwof\eta^{CD}\tetradf_B{}^{[\mu}\tetradf_C{}^{\nu]}\tetradf_D{}^{[\rho}\tetradf_A{}^{\sigma]}-\constthreef\eta^{CD}\tetradf_A{}^{[\mu}\tetradf_C{}^{\nu]}\tetradf_D{}^{[\rho}\tetradf_B{}^{\sigma]}
	\end{split},
	\end{align}
	where we introduced, a priory, arbitrary coefficients $\constonef,\consttwof,\constthreef$, for the part which only depends on $\ff$ and $\constonekin,\consttwokin,\constthreekin$ for the terms with derivative interactions. For the linear analysis, we will use a different supermetric
	\begin{align}
	\accentset{kin}{G}_{AB}{}^{\mu\nu\rho\sigma}=
	\constonekin
	\gee_{\alpha\delta}\, \gee^{\kappa\beta}\, \gf^{\lambda\gamma} \tetrade_A{}^\alpha \tetradf_B{}^\delta \,
	+\consttwokin
	\gee^{\kappa\beta}\, \tetrade_A{}^\delta \tetradf_B{}^\gamma \,
	+\constthreekin
	\gee^{\kappa\beta}\, \tetrade_A{}^\alpha \tetradf_B{}^\gamma\, ,
	\end{align}
	pointing out that its use in our perturbation theory still gives compatible results since we only look at lower order expansions around a Minkowski background, where the contractions are reduced to contractions with the Minkowski metric.

	\section{Linear analysis of teleparallel bigravity}
	\label{sec:Linear}
	In this section we examine teleparallel bigravity at the linear level, where we retain only linear perturbations of the field equations or, equivalently, second order perturbations of the Lagrangian. The results motivate a further examination of these theories at the nonlinear level, which we carry out in section \ref{sec:Nonlinear}.
	\subsection{The spectrum of teleparallel bigravity}
	\label{sec:TBspectrum} The full action and Lagrangian density of the theory are specified as
	\begin{align} \label{eq:8glag}
	S=\int\td^4 x \, \left(\overset{\ee}\LG+\overset{\ff}\LG+\overset{kin}\LG\right)-\overset{int}\LG \,,
	\end{align}
	where
	\begin{align}
	\overset{\ee}\LG:&= -\detg\,m^2_\ee\Big(
	\constonee \gee_{\alpha\delta}\, \gee^{\kappa\beta}\, \gee^{\lambda\gamma} \tetrade_A{}^\alpha \tetrade_B{}^\delta \,
	\accentset{\ee}{T}^A{}_{\kappa \lambda}\,\accentset{\ee}{T}^B{}_{\beta \gamma}
	+\consttwoe
	\gee^{\kappa\beta}\, \tetrade_A{}^\delta \tetrade_B{}^\gamma \,
	\accentset{\ee}{T}^A{}_{\kappa \gamma}\,\accentset{\ee}{T}^B{}_{\beta \delta}
	+\constthreee
	\gee^{\kappa\beta}\, \tetrade_A{}^\alpha \tetrade_B{}^\gamma\, \accentset{\ee}{T}^A{}_{\kappa \alpha} \,\accentset{\ee}{T}^B{}_{\beta \gamma}
	\Big)
	\\
	\overset{\ff}\LG:&= -\detf\,m^2_\ff\Big(
	\constonef
	\gf_{\alpha\delta}\, \gf^{\kappa\beta}\, \gf^{\lambda\gamma} \tetradf_A{}^\alpha \tetradf_B{}^\delta \,
	\accentset{\ff}{T}^A{}_{\kappa \lambda}\,\accentset{\ff}{T}^B{}_{\beta \gamma}
	+\consttwof
	\gee^{\kappa\beta}\, \tetradf_A{}^\delta \tetradf_B{}^\gamma \,
	\accentset{\ff}{T}^A{}_{\kappa \gamma}\,\accentset{\ff}{T}^B{}_{\beta \delta}
	+\constthreef
	\gee^{\kappa\beta}\, \tetradf_A{}^\alpha \tetradf_B{}^\gamma\, \accentset{\ff}{T}^A{}_{\kappa \alpha} \,\accentset{\ff}{T}^B{}_{\beta \gamma}
	\Big)
	\\
	\overset{kin}\LG&:=-\detg\,m^2_{\ee\ff}\Big(
	\constonekin
	\gee_{\alpha\delta}\, \gee^{\kappa\beta}\, \gf^{\lambda\gamma} \tetrade_A{}^\alpha \tetradf_B{}^\delta \,
	\accentset{\ee}{T}^A{}_{\kappa \lambda}\,\accentset{\ff}{T}^B{}_{\beta \gamma}
	+\consttwokin
	\gee^{\kappa\beta}\, \tetrade_A{}^\delta \tetradf_B{}^\gamma \,
	\accentset{\ee}{T}^A{}_{\kappa \gamma}\,\accentset{\ff}{T}^B{}_{\beta \delta}
	+\constthreekin
	\gee^{\kappa\beta}\, \tetrade_A{}^\alpha \tetradf_B{}^\gamma\, \accentset{\ee}{T}^A{}_{\kappa \alpha} \,\accentset{\ff}{T}^B{}_{\beta \gamma}
	\Big)
	\\
	\begin{split}\overset{int}\LG &:=m^4_{int}\epsilon_{ABCD}
	\Big(
	\beta_0 E^A\wedge E^B\wedge E^C\wedge E^D
	+
	\beta_1 E^A\wedge E^B\wedge E^C\wedge F^D
	+
	\beta_2E^A\wedge E^B\wedge F^C\wedge F^D
	\\
	&	+
	\beta_3E^A\wedge F^B\wedge F^C\wedge F^D
	+
	\beta_4F^A\wedge F^B\wedge F^C\wedge F^D
	\Big)
	\end{split}
	\end{align}
	The expressions used for the perturbations of the tetrad, cotetrad and volume form are
	\begin{align}
	{\ee^A}_\delta
	&\to
	{\delta^A}_\delta + \frac\epsilon 2 \,\eta^{AB} {\delta_B}^\alpha \left(
	\he_{\alpha\delta} + \be_{\alpha\delta}\right)
	=
	{\delta^A}_\delta + \epsilon \, \eta^{AB} {\delta_B}^\alpha \Be_{\alpha\delta};
	\qquad \Be_{\alpha\delta}:=\frac{\he_{\alpha\delta} + \be_{\alpha\delta}}{2}
	\\
	{\tetrade_A}^\gamma
	&\to
	{\delta_A}^\gamma- \gbe^{\gamma\beta}{\delta_A}^\alpha\left(\frac{\epsilon}{2}\left(\he_{\beta\alpha}+\be_{\beta\alpha}\right)
	+\frac{\epsilon^2}{4}\left(\be_{\alpha\nu}{\be_\beta}^\nu-\he_{\alpha\nu}{\be_\beta}^\nu+\be_{\alpha\nu}{\he_\beta}^\nu-\he_{\alpha\nu}{\he_\beta}^\nu\right)\right)
	= \\ \nn
	&\quad =
	{\delta_A}^\gamma- \gbe^{\gamma\beta}{\delta_A}^\alpha\left(\epsilon\,\frac{\he_{\beta\alpha}+\be_{\beta\alpha}}{2}
	+\epsilon^2\,\frac{-{\be_\beta}^\nu\be_{\nu\alpha}-{\be_\beta}^\nu\he_{\nu\alpha}-{\he_\beta}^\nu\be_{\nu\alpha}-{\he_\beta}^\nu\he_{\nu\alpha}}{4}\right)=\\ \nn
	&\quad =
	{\delta_A}^\gamma+ \gbe^{\gamma\beta}{\delta_A}^\alpha\left(-\epsilon\,\Be_{\beta\alpha}
	+\epsilon^2\,\Be_\beta{}^\nu \Be_{\nu\alpha}\right),
	\end{align}
	\begin{align}
		\label{eq:deteexp}
		\detg
		\equiv
		\frac{1}{4!}\epsilon_{ABCD}\epsilon^{\alpha \beta \gamma \delta}{\ee^A}_\alpha{\ee^B}_\beta{\ee^C}_\gamma{\ee^D}_\delta
		\to 
		\bar \ee\left(1 + \frac12 {\he^A}_A +\frac18 \be_{AB} \be^{AB}-\frac18 \he_{AB} \he^{AB}+\frac18 ({\he^A}_A)^2\right)
		\end{align}
	where we used the permutation symbol $\epsilon_{ABCD}$. Analogous relations hold for the $\tetradf$ tetrad. Notice the presence of the background metric $ \gbe^{\gamma\beta}$ and that $\bar \ee = \sqrt{\modu{\gbe}}$.

	The resulting linearized Lagrangian then is
	\begin{align}
	\label{eq:lalag}
	S=\int\td^4x\left(
	\overset{\ee}\LG_{hh} + \overset{\ee}\LG_{hb} + \overset{\ee}\LG_{bb}
	+
	\overset{\ff}\LG_{hh} + \overset{\ff}\LG_{hb} + \overset{\ff}\LG_{bb}
	+
	\overset{kin}\LG_{hh} + \overset{kin}\LG_{hb} + \overset{kin}\LG_{bb}
	+
	\overset{int}\LG_{hh} + \overset{int}\LG_{hb} + \overset{int}\LG_{bb}
	+
	\overset{int}\LG_{1} + \overset{int}\LG_{0}
	\right)
	\end{align}
	with the second order expansions given by:
	\begin{align}
	\overset{\ee}\LG_{hh} &=\frac{m_\ee^2}{4}\he^{AB}
	\left(
	-2 \,\constthreee  \pd_B\pd_A \he
	+(-2 \constonee-\consttwoe+\constthreee)\pd_C\pd_B\he_A{}^C
	+(2\constonee+\consttwoe)\square\he_{AB}
	+\constthreee\eta_{AB}\,\square\he
	\right)
	\\
	\overset{\ee}\LG_{hb} &=-\frac{m_\ee^2}{2}\he^{AB}
	\left(2 \constonee+\consttwoe+\constthreee\right)\pd_C\pd_B\be_A{}^C
	\\
	\overset{\ee}\LG_{bb} &=\frac{m_\ee^2}{4}\be^{AB}
	\left((-2\constonee+3\consttwoe+\constthreee)\pd_C\pd_B\be_A{}^C+(2\constonee-\consttwoe)\square \be_{AB}
	\right)
	\\ \nonumber
	\\\nn
	\end{align}
	\begin{align}
	\overset{\ff}\LG_{hh} &=\frac{m_\ff^2}{4}\hf^{AB}
	\left(
	-2 \constthreef \pd_B\pd_A \hf
	+(-2 \constonef-\consttwof+\constthreef)\pd_C\pd_B\hf_A{}^C
	+(2\constonef+\consttwof)\square\hf_{AB}
	+\constthreef\eta_{AB}\,\square\hf
	\right)
	\\
	\overset{\ff}\LG_{hb} &=-\frac{m_\ff^2}{2}\hf^{AB}
	\left(2 \constonef+\consttwof+\constthreef\right)\pd_C\pd_B\bff_A{}^C
	\\
	\overset{\ff}\LG_{bb} &=\frac{m_\ff^2}{4}\bff^{AB}
	\left((-2\constonef+3\consttwof+\constthreef)\pd_C\pd_B\bff_A{}^C+(2\constonef-\consttwof)\square \bff_{AB}
	\right)
	\\ \nonumber
	\\\nn
	\end{align}
	\begin{align}
	\overset{kin}\LG_{hh} &=\frac{m_{\ee\ff}^2}{4}\he^{AB}
	\left(
	- \constthreekin \pd_B\pd_A \hf
	+(-2 \constonekin-\consttwokin+\constthreekin)\pd_C\pd_B\hf_A{}^C
	+(2\constonekin+\consttwokin)\square\hf_{AB}
	+\constthreekin\eta_{AB}\,(-\pd_C\pd_B\hf^{BC}+\square\hf)
	\right)
	\\
	\overset{kin}\LG_{hb} &=-\frac{m_{\ee\ff}^2}{4}
	\left(2\constonekin+\consttwokin+\constthreekin\right)
	\left(\he^{AB}\pd_C\pd_B\bff_A{}^C + \be^{AB}\pd_C\pd_B\hf_A{}^C\right)
	\\
	\overset{kin}\LG_{bb} &=\frac{m_{\ee\ff}^2}{4}\be^{AB}
	\left((-2\constonekin+3\consttwokin+\constthreekin)\pd_C\pd_B\bff_A{}^C + (2\constonekin-\consttwokin)\square \bff_{AB}
	\right)
	\\ \nonumber
	\\\nn
	\end{align}
	\begin{align}
	\overset{int}{\LG}_0 &= -(\beta_0+\beta_1+\beta_2+\beta_3+\beta_4) m^4_{int}
	\\
	\overset{int}{\LG}_1 &=
	-\frac{ m^4_{int}}{8}\left((4\beta_0 + 3\beta_1 + 2\beta_2 + \beta_3) \he
	+ (\beta_1 + 2\beta_2 + 3\beta_3 + 4\beta_4)\hf
	\right)
	\\
	\overset{int}{\LG}_{hh} & =\frac{ m^4_{int}}{48}
	\left((6\beta_0 + 3\beta_1 + \beta_2)\he_{AB}\he^{AB}+ (\beta_2 + 3\beta_3+6\beta_4)\hf^{AB}\hf_{AB}
	+ (3\beta_1 + 4\beta_2 +3\beta_3 )\he^{AB}\hf_{AB}+
	\right.\\ \nn
	&\quad\left.
	-(6\beta_0 + 3\beta_1 + \beta_2)\he\he - (\beta_2 + 3\beta_3 + 6\beta_4) \hf\hf -(3\beta_1+4\beta_2+3\beta_3)\he\hf
	\right)
	\\
	\overset{int}{\LG}_{hb} & = 0
	\\
	\overset{int}{\LG}_{bb} & = -\frac{m^4_{int}}{48} \left(
	(6\beta_0 + 3 \beta_1 + \beta_2)\be_{AB}\be^{AB} + (\beta_2 + 3\beta_3 + 6\beta_4) \bff_{AB}\bff^{AB}
	+ (3\beta_1 + 4\beta_2 + 3\beta_3)\bff_{AB}\be^{AB}
	\right)\,,
	\end{align}
	where we defined $\he:=\he^A{}_A$ and $\hf:=\hf^A{}_A$.

	\subsection{Linear analysis of the teleparallel equivalent to Hassan-Rosen bimetric gravity}

	As a first application of the general theory, we recover the standard bimetric theory in tetrad formulation. To this purpose we eliminate the new terms brought by the kinetic mixing in the Lagrangian setting $\accentset{\textrm{\ae}}{C}_i=0$. For the sake of visualization, we also rewrite the above Lagrangian in matrix formalism, using a basis where the field vector has the components $\Phi:=$($\he_{AB}$, $\hf_{AB}$, $\be_{AB}$, $\bff_{AB}$)$^T$ and by integrating by parts if required
	\begin{align}
	\LG =\Phi^T(\mathbb{K}+\mathbb{M}^2)\Phi+\overset{int}{\LG}_1+\overset{int}{\LG}_0.
	\end{align}
	The matrices $\mathbb{K}$ and $\mathbb{M}^2$ for the general case can be straightforwardly read from the Lagrangian in Eq.~\eqref{eq:lalag}. Below we give their explicit forms only for the cases under consideration.

	Requiring the absence of linear terms for $\he$ and $\hf$ in the bimetric Lagrangian yields the following conditions on the coefficients entering the interaction term
	$$
	\begin{cases}
	4\beta_0 = - 3\beta_1 - 2 \beta_2 -\beta_3\\
	4\beta_4 = - 3\beta_3 - 2 \beta_2 -\beta_1
	\end{cases}
	$$
	and, as a result, we also obtain that
	$$
	\beta_0 + \beta_1 + \beta_2 + \beta_3 + \beta_4 \equiv 0,
	$$
	in the same fashion, for instance, as the condition (3.2) and (3.3) of \cite{Hinterbichler:2012cn}. After the cancellation of these tadpole terms, enforcing $\overset{int}{\LG}_1=\overset{int}{\LG}_0=0$, the mass term reduces to
	\begin{align}
	\mathbb{M}^2&:=\dfrac{ m^4_{int}\left(3\beta_1 + 4\beta_2 +3\beta_3 \right)}{96}
	\begin{pmatrix}
	-1 & 1 & 0 & 0\\
	1 & -1 & 0 & 0\\
	0 & 0 & 1 & -1\\
	0 & 0 & -1 & 1
	\end{pmatrix}
	\left(\eta^{AD}\eta^{BE}-\eta^{AB}\eta^{DE}\right),
	\end{align}
	where, in the last step, we used  $\bff_A{}^A=\be_A{}^A\equiv0$.

	As for the kinetic terms, in order to recover standard bimetric theory, we enforce the cancellation of the $h-b$ mode mixing by setting $-2\constonee=\consttwoe+\constthreee$, as well as $-2\constonef=\consttwof+\constthreef$ yielding two copies of the one parameter family of new general relativity introduced in \cite{PhysRevD.19.3524}. The resulting kinetic term presents then the following structure
	\begin{align}
	\mathbb{K}&:=
	-	\frac14
	\begin{pmatrix}
	m_\ee^2 \,\constthreee
	&
	0
	&
	0
	&
	0
	\\
	0
	&
	m_\ff^2 \,\constthreef
	&
	0
	&
	0
	\\
	0
	&
	0
	&
	m_\ee^2(2\consttwoe+\constthreee)
	&
	0
	\\
	0
	&
	0
	&
	0
	&
	m_\ff^2(2\consttwof+\constthreef)
	\end{pmatrix}
	\mathcal{E}^{ABDE}
	\end{align}
	where the \textit{Lichnerowicz operator} is defined as
	\begin{align}
	\mathcal{E}^{ABDE}:=\eta^{AD}\eta^{BE}\square-2\eta^{AD}\pd^E\pd^B+(\eta^{DE}\pd^A\pd^B+\eta^{AB}\pd^D\pd^E)-\eta^{AB}\eta^{DE}\square
	\end{align}
	and we have again used the fact that $\bff_A{}^A=\be_A{}^A\equiv0$.

	With the kinetic term being already in a diagonal form, we need only to diagonalize the remaining mass term. As a first step, we give the fields the canonical mass dimension through the field redefinition
	\begin{align}
	\he_{AB}\to\frac{\he_{AB}}{m_\ee}, \quad \hf_{AB}\to\frac{\hf_{AB}}{m_\ff}, \quad
	\be_{AB}\to\frac{\be_{AB}}{m_\ee}, \quad \bff_{AB}\to\frac{\bff_{AB}}{m_\ff}
	\end{align}
	which puts the Lagrangian in the form:
	\begin{align}
	\begin{split}
	\LG_{bm} &= \frac14
	\begin{pmatrix}
	\he_{AB}& \hf_{AB} & \be_{AB} & \bff_{AB}
	\end{pmatrix}
	\left(
	\mathbb{K}_{bm}
	+\mathbb{M}^2_{bm}
	\right)
	\begin{pmatrix}
	\he_{DE}\\ \hf_{DE} \\ \be_{DE} \\ \bff_{DE}
	\end{pmatrix}\,.
	\end{split}
	\end{align}
	with
	\begin{align}
	\mathbb{K}_{bm}=
	-\begin{pmatrix}
	\,\constthreee
	&
	0
	&
	0
	&
	0
	\\
	0
	&
	\,\constthreef
	&
	0
	&
	0
	\\
	0
	&
	0
	&
	2\consttwoe+\constthreee
	&
	0
	\\
	0
	&
	0
	&
	0
	&
	2\consttwof+\constthreef
	\end{pmatrix}
	\mathcal{E}^{ABDE},
	\end{align}
	and
	\begin{align}
	\mathbb{M}^2_{bm}=\dfrac{ m^4_{int}\left(3\beta_1 + 4\beta_2 +3\beta_3 \right)}{24}
	\begin{pmatrix}
	-\dfrac{1}{m_\ee^2} & \dfrac{1}{m_\ee m_\ff} & 0 & 0\\
	\dfrac{1}{m_\ee m_\ff} & -\dfrac{1}{m_\ff^2} & 0 & 0\\
	0 & 0 & \dfrac{1}{m_\ee^2 } & -\dfrac{1}{m_\ee m_\ff}\\
	0 & 0 & -\dfrac{1}{m_\ee m_\ff} & \dfrac{1}{m_\ff^2}
	\end{pmatrix}
	\left(\eta^{AD}\eta^{BE}-\eta^{AB}\eta^{DE}\right).
	\end{align}
	Before analyzing the two emerging blocks in isolation, we further simplify the Lagrangian by writing, without loss of generality,
	\begin{align}
	\label{eq:mfp}
	\dfrac{ m^4_{int}\left(3\beta_1 + 4\beta_2 +3\beta_3 \right)}{24}=:m_{FP}^2\frac{m_\ee^2 m_\ff^2}{m_\ee^2+ m_\ff^2}\,.
	\end{align}
	As usual in bimetric theories, we have re-absorbed the dependence of the Lagrangian on $\beta_{i=1,2,3}$ into the definition of the mass scale $m_{FP}$.

	\subsubsection{Symmetric sector}

	Focusing now on the symmetric perturbations, we have
	\begin{align}
	\begin{split}
	\LG_{bm}^{hh} &= -\frac{1}{4}\he_{AB} \, \constthreee \,\mathcal{E}^{ABDE}\,\he_{DE} -\frac{1}{4}\hf_{AB} \, \constthreef \,\mathcal{E}^{ABDE}\,\hf_{DE}\\
	&-\frac{m_{FP}^2}{4}\frac{m_\ee^2 m_\ff^2}{m_\ee^2+ m_\ff^2}
	\left[\left(\frac{\he_{AB}}{m_\ee}-\frac{\hf_{AB}}{m_\ff}\right)\left(\frac{\he^{AB}}{m_\ee}-\frac{\hf^{AB}}{m_\ff}\right)
	-
	\left(\frac{\he}{m_\ee}-\frac{\hf}{m_\ff}\right)^2
	\right]
	\end{split}
	\end{align}
	which is diagonalized through the usual field redefinition
	\begin{align}
	M_{AB}:=\frac{m_\ee m_\ff}{\sqrt{m_\ee^2 + m_\ff^2}}\left(\frac{\he_{AB}}{m_\ee}-\frac{\hf_{AB}}{m_\ff}\right)
	=\frac{m_\ff }{\sqrt{m_\ee^2 + m_\ff^2}}\he_{AB} - \frac{m_\ee }{\sqrt{m_\ee^2 + m_\ff^2}}\hf_{AB}
	\end{align}
	ensuring that the norm of $M_{AB}$ in field space is the unity.

	The orthogonal combination $G_{AB} = \alpha \he_{AB} + \beta \hf_{AB}$ is determined through
	\begin{align}
	\begin{cases}
	G_{AB}\cdot M_{AB} = 0
	\\
	G_{AB}\cdot G_{AB} = M_{AB}\cdot M_{AB}
	\end{cases}
	\implies
	\begin{cases}
	\dfrac{\alpha}{m_\ee} - \dfrac{\beta}{m_\ff}=0
	\\
	\alpha^2 + \beta^2 = 1
	\end{cases}
	\implies
	\begin{cases}
	\alpha = \dfrac{m_\ee\beta}{m_\ff}
	\\
	\beta^2\left(1+\dfrac{m_\ee^2}{m_\ff^2} \right) = 1
	\end{cases}
	\implies
	\begin{cases}
	\alpha = \dfrac{m_\ee}{\sqrt{m_\ee^2+m_\ff^2}}
	\\
	\beta= \dfrac{m_\ff}{\sqrt{m_\ee^2+m_\ff^2}}
	\end{cases}
	\end{align}
	where the $\cdot$ operator stands for the scalar product in field space. As a result, we obtain
	\begin{align}
	G_{AB}:=\frac{m_\ee }{\sqrt{m_\ee^2 + m_\ff^2}}\he_{AB}+\frac{ m_\ff}{\sqrt{m_\ee^2 + m_\ff^2}}\hf_{AB}\,,
	\end{align}
	showing that the field transformation linking the mass eigenstates to the original fields is simply a rotation in field space:
	\begin{align}
	\begin{pmatrix}
	\label{eq:eigrot}
	M_{AB}\\G_{AB}
	\end{pmatrix}
	=
	\begin{pmatrix}
	\dfrac{m_\ff }{\sqrt{m_\ee^2 + m_\ff^2}} & -\dfrac{m_\ee }{\sqrt{m_\ee^2 + m_\ff^2}}\\
	\dfrac{m_\ee }{\sqrt{m_\ee^2 + m_\ff^2}} & \dfrac{m_\ff }{\sqrt{m_\ee^2 + m_\ff^2}}
	\end{pmatrix}
	\begin{pmatrix}
	\he_{AB}\\ \hf_{AB}
	\end{pmatrix}\,.
	\end{align}
	In terms of the mass eigenstates, the Lagrangian of the symmetric sector is then given by 
	\begin{align}
	\LG_{bm}^{hh} &=
	\frac{1}{4}G_{AB} \, \frac{\constthreee m_\ee^2 + \constthreef m_\ff^2}{m_\ee^2 + m_\ff^2} \,\tilde{\mathcal{E}}^{ABDE}\,G_{DE}
	+
	\frac{1}{4} M_{AB} \, \frac{\constthreef m _\ee^2+ \constthreee m_\ff^2}{m_\ee^2 + m_\ff^2} \,\tilde{\mathcal{E}}^{ABDE}\,M_{DE}
	\\\nn&+
	\frac{1}{2} M_{AB} \, \frac{\left(\constthreee -\constthreef\right) m _\ee m_\ff}{m_\ee^2 + m_\ff^2} \,\tilde{\mathcal{E}}^{ABDE}\,G_{DE}
	-
	\frac{m_{FP}^2}{4}
	\left(M_{AB}M^{AB} - M^2 \right)\,,
	\end{align}
	showing that the condition $\constthreee =\constthreef$ is necessary for a complete diagonalization of the symmetric sector. In fact, because the performed field redefinition is simply a rotation, the kinetic term matrix of the original field must be proportional to the unit matrix in order to be unaffected by the field redefinition.

	\subsubsection{Antisymmetric sector}

	As for the antisymmetric sector, we have
	\begin{align}
	\begin{split}
	\LG_{bm}^{bb} &= -\frac{1}{4}\be_{AB} \,\left(2\consttwoe +\constthreee\right) \,\mathcal{E}^{ABDE}\,\be_{DE} -\frac{1}{4}\bff_{AB} \, \left(2\consttwof+\constthreef\right) \,\mathcal{E}^{ABDE}\,\bff_{DE}\\
	&+\frac{m_{FP}^2}{4}\frac{m_\ee^2 m_\ff^2}{m_\ee^2+ m_\ff^2}
	\left(\frac{\be_{AB}}{m_\ee}-\frac{\bff_{AB}}{m_\ff}\right)\left(\frac{\be^{AB}}{m_\ee}-\frac{\bff^{AB}}{m_\ff}\right)
	\end{split}
	\end{align}
	which is diagonalized through a rotation analogous to that of the symmetric sector. In particular, by defining
	\begin{align}
	\label{eq:EFsub}
	\begin{pmatrix}
	E_{AB} \\
	F_{AB}
	\end{pmatrix}
	=
	\begin{pmatrix}
	\dfrac{m_\ff }{\sqrt{m_\ee^2 + m_\ff^2}} & -\dfrac{m_\ee }{\sqrt{m_\ee^2 + m_\ff^2}}\\
	\dfrac{m_\ee }{\sqrt{m_\ee^2 + m_\ff^2}} & \dfrac{m_\ff }{\sqrt{m_\ee^2 + m_\ff^2}}
	\end{pmatrix}
	\begin{pmatrix}
	\be_{AB}\\ \bff_{AB}
	\end{pmatrix}
	\end{align}
	we have
	\begin{align}
	\LG_{bm}^{bb} &=
	-\frac{1}{4} F_{AB} \, \frac{\left(2\consttwoe+\constthreee\right) m_\ee^2 + \left(2\consttwof+\constthreef\right) m_\ff^2}{m_\ee^2 + m_\ff^2} \,\mathcal{E}^{ABDE}\,F_{DE} \\\nn&
	-
	\frac{1}{4}E_{AB} \, \frac{\left(2\consttwoe+\constthreee\right) m_\ff^2 + \left(2\consttwof+\constthreef\right) m_\ee^2}{m_\ee^2 + m_\ff^2} \,\mathcal{E}^{ABDE}\,E_{DE}
	\\\nn&
	-
	\frac12
	F_{AB} \, \frac{\left(2\consttwoe+\constthreee -2\consttwof-\constthreef\right) m _\ee m_\ff}{m_\ee^2 + m_\ff^2} \,\mathcal{E}^{ABDE}\,E_{DE}+\frac{m_{FP}^2}{4}E_{AB}E^{AB}\,.
	\end{align}
	In analogy with the case of the symmetric sector, the condition $2\consttwoe + \constthreee = 2\consttwof + \constthreef  $ ensures the decoupling of the modes.

	\subsubsection{Teleparallel Equivalent of Bimetric Gravity}

	In terms of the mass eigenstates we have identified and through the above conditions, the full Lagrangian then becomes:
	\begin{align}
	\LG_{bm} &=
	-\frac{1}{4} \constthreee \, G_{AB} \,\mathcal{E}^{ABDE}\,G_{DE}
	-
	\frac{1}{4} \constthreee \, M_{AB} \,\mathcal{E}^{ABDE}\,M_{DE}
	-
	\frac{m_{FP}^2}{4}
	\left(M_{AB}M^{AB} - M^2 \right)
	\\\nn& \quad
	-\frac{1}{4} \left(2\consttwoe+\constthreee\right)\, F_{AB} \,\mathcal{E}^{ABDE}\,F_{DE}
	-
	\frac{1}{4}\left(2\consttwoe+\constthreee\right) \, E_{AB} \,\mathcal{E}^{ABDE}\,E_{DE}
	+\frac{m_{FP}^2}{4}E_{AB}E^{AB}\,.
	\end{align}
	Setting now the remaining coefficients to their TEGR values, we notice that $\constthreee=\constthreef=-1$ ensures a proper normalizes the Lichnerowicz operator $\mathcal{E}^{ABDE}$ and casts the Lagrangian of the symmetric sector into the usual bimetric form:
	\begin{align}
	\label{eq:GMlag}
	\LG_{bm}^{hh} =
	\frac{1}{4}G_{AB} \, \mathcal{E}^{ABDE}\,G_{DE}
	+
	\frac{1}{4} M_{AB} \, \mathcal{E}^{ABDE}\,M_{DE}
	-
	\frac{m_{FP}^2}{4}
	\left(M_{AB}M^{AB} - M^2 \right)
	\,.
	\end{align}
	Differently, the choice $\consttwoe=\consttwof=C_2=1/2$ prevents the propagation of all antisymmetric modes, reducing the corresponding Lagrangian to
	\begin{align}
	\LG_{bm}^{bb}=\frac{m_{FP}^2}{4}E_{AB}E^{AB}\,.
	\end{align}
	The resulting equation of motion then forces constraint $E_{AB} = 0$, which through eq.$\eqref{eq:EFsub}$ recovers
	\begin{align}
	\frac{\be_{AB}}{m_\ee} = \frac{\bff_{AB}}{m_\ff}\,.
	\end{align}
	At the first order of the perturbation expansion, this condition coincides with the requirement  $\eta_{AB}{\ee^A}_{[\mu} {\ff^B}_{\nu]}\req0$, often discussed in the literature of bigravity \cite{Kocic:2018ddp,Kocic:2018yvr,Kocic:2019ahm,Hinterbichler:2012cn}.

	We remark that the final form of the Lagrangian was obtained by using a rotation in field space, rather than a generic linear transformation, to diagonalize the mass term. Whether this might seem as an overly restrictive choice, we remind that in the teleparallel equivalent of bimetric gravity, the coefficients $C_i$ of the two tetrads are set from the very beginning to their TEGR values and, as a consequence, the identified rotation is the only transformation that recasts the symmetric sector in the standard form.

	We also notice that prior to setting the coefficients $C_i$ to their TEGR values, the Lagrangian of the antisymmetric sector mirrors that of the symmetric one, albeit an important distinction. The mass terms of the massive modes, in fact, have opposite signs. Whether a generic theory containing propagating symmetric and antisymmetric modes would probably contain ghosts, at this stage it seems still possible to recover a regime of the theory characterized by the propagation of a massive and a massless antisymmetric mode and no symmetric perturbations.

	\subsection{Linear analysis of the full Lagrangian}

We now face the full Lagrangian, allowing for the kinetic mixing terms that we neglected in the previous section. Note that in the limit \(m_{FP} \to 0\) of vanishing Fierz-Pauli mass, a kinetic coupling in the symmetric sector, which contains two massless spin-2 fields in this case, inevitably leads to the existence of a scalar ghost mode~\cite{Boulanger:2000rq}. For this reason, such kinetic coupling is not considered in the Hassan-Rosen bigravity~\cite{Hassan:2011zd}.

	The treatment of the linear and 0-th order terms proceeds as in the standard bimetric case: enforcing the absence of tadpole term by setting
	\begin{align}
	\begin{cases}
	4\beta_0 = - 3\beta_1 - 2 \beta_2 -\beta_3\\
	4\beta_4 = - 3\beta_3 - 2 \beta_2 -\beta_1
	\end{cases}\,,
	\end{align}
	we obtain $\beta_0 + \beta_1 + \beta_2 + \beta_3 + \beta_4 \equiv 0$, and the mass term consequently acquires again the form
	\begin{align}
	\mathbb{M}^2=
	\dfrac{ m^4_{int}\left(3\beta_1 + 4\beta_2 +3\beta_3 \right)}{96}
	\begin{pmatrix}
	-\dfrac{1}{m_\ee^2} & \dfrac{1}{m_\ee m_\ff} & 0 & 0\\
	\dfrac{1}{m_\ee m_\ff} & -\dfrac{1}{m_\ff^2} & 0 & 0\\
	0 & 0 & \dfrac{1}{m_\ee^2 } & -\dfrac{1}{m_\ee m_\ff}\\
	0 & 0 & -\dfrac{1}{m_\ee m_\ff} & \dfrac{1}{m_\ff^2}
	\end{pmatrix}
	\left(\eta^{AD}\eta^{BE}-\eta^{AB}\eta^{DE}\right)
	\end{align}
	in terms of the properly normalized fields given by
	\begin{align}
	\he_{AB}\to\frac{\he_{AB}}{m_\ee}, \quad \hf_{AB}\to\frac{\hf_{AB}}{m_\ff}, \quad
	\be_{AB}\to\frac{\be_{AB}}{m_\ee}, \quad \bff_{AB}\to\frac{\bff_{AB}}{m_\ff}\,.
	\end{align}
	As for the kinetic term of the normalized fields, in this first analysis we choose to set
	\begin{align}
	\begin{cases}
	-2\constonee = \consttwoe + \constthreee\\
	-2\constonef = \consttwof + \constthreef\\
	-2\constonekin = \consttwokin + \constthreekin
	\end{cases}
	\end{align}
	in order to obtain the same block diagonal structure of the mass term. Whereas this condition is in principle unnecessary, it allows us to tackle the symmetric and antisymmetric modes separately, with a considerable simplification of the diagonalization procedure. In order to further simplify the Lagrangian, we also opt to set $m^2_{\ee\ff}\req m_\ee m_\ff$, obtaining:
	\begin{align}
	\mathbb{K}&=
	-\frac14
	\begin{pmatrix}
	\constthreee & \dfrac{\constthreekin}{2} & 0 &0\\
	\dfrac{\constthreekin}{2}& \constthreef & 0&0\\
	0&0&(2\consttwoe+\constthreee)& \dfrac{1}{2}
	(2\consttwokin+\constthreekin)\\
	0&0& \dfrac{1}{2}
	(2\consttwokin+\constthreekin)&(2\consttwof+\constthreef)
	\end{pmatrix}
	\mathcal{E}^{ABDE}\,,
	\end{align}
	with the Lichnerowicz operator $\mathcal{E}^{ABDE}$ defined as in the standard case. Because the interaction term is the same as in the standard bimetric gravity, the field rotations in eq.$\eqref{eq:eigrot}$ and $\eqref{eq:EFsub}$ certainly diagonalize it. However, if we restrict ourselves to these transformations,  the coefficients in the kinetic term will have to be set in a way that the kinetic term becomes proportional to the unit matrix in field space. In particular, this choice will force the absence of kinetic mixing, leading inevitably to scenarios close to the standard bimetric case or that, at most, allow for the propagation of a massless and a negative mass antisymmetric mode. For this reason, in the following we consider instead generic superpositions of symmetric and antisymmetric perturbations rather than rotations in the corresponding field spaces.

	\subsubsection{Symmetric sector}

	Upon partial integration of the off-diagonal kinetic terms, the Lagrangian for the symmetric perturbation is written as:
	\begin{align}
	\LG^{hh}
	&=
	-\frac14\,\he_{AB} \constthreee\mathcal{E}^{ABDE}\,\he_{DE}
	-
	\frac14\,\hf_{AB} \constthreef\mathcal{E}^{ABDE}\,\hf_{DE}
	-
	\frac14\,\he_{AB} \constthreekin\mathcal{E}^{ABDE}\,\hf_{DE}
	\\ \nn
	&
	-\dfrac{ m^4_{int}}{4}
	\left[\left(\frac{\he_{AB}}{m_\ee}-\frac{\hf_{AB}}{m_\ff}\right)\left(\frac{\he^{AB}}{m_\ee}-\frac{\hf^{AB}}{m_\ff}\right)
	-
	\left(\frac{\he}{m_\ee}-\frac{\hf}{m_\ff}\right)^2
	\right]
	\end{align}
	where, without loss of generality, we have set $\left(3\beta_1 + 4\beta_2 +3\beta_3 \right)=24$. In order to diagonalize the Lagrangian, we now consider the generic linear transformation
	\begin{align}
	\label{eq:gen_t}
	\begin{pmatrix}
	\he_{AB}\\\hf_{AB}
	\end{pmatrix}
	=:
	\begin{pmatrix}
	\alpha& \beta \\
	\gamma &\delta
	\end{pmatrix}
	\begin{pmatrix}
	M_{AB}\\G_{AB}
	\end{pmatrix}
	\end{align}
	where $G_{AB}$ and $M_{AB}$ are respectively the graviton and massive bigraviton excitation. However generic the transformation be, we still require that an inverse exists by imposing that \textit{the determinant of the transformation never vanishes}. We will verify that the solutions we identify indeed satisfy this criterion.

	Analyzing the mass term in isolation, we see that
	\begin{align}
	\left(\frac{\he_{AB}}{m_\ee}-\frac{\hf_{AB}}{m_\ff}\right) = \left[\left(\frac{\alpha}{m_\ee}-\frac{\gamma}{m_\ff}\right)\,M_{AB}+\left(\frac{\beta}{m_\ee}-\frac{\delta}{m_\ff}\right)\,G_{AB}\right]
	\end{align}
	and analogous expressions hold for the remaining terms. It is then clear that setting
	\begin{align}
	\label{eq:noGmass}\delta=\frac{m_\ff}{m_\ee}\beta
	\end{align}
	forbids a mass term for the $G$ perturbation. Similarly, requiring that the kinetic terms for $G$ and $M$ have the canonical form, as well as the presence of no mixing, results in three further conditions that fully determine the parameters in eq.$\eqref{eq:gen_t}$. Explicitly, we have
	\begin{align}
	\begin{split}
	\LG^{hh} &\supset
	-\frac{1}{4}\left(\constthreee \alpha^2 + \constthreef \gamma^2 + \constthreekin\alpha\gamma \right)
	\,M_{AB} \,\mathcal{E}^{ABDE}\,M_{DE}
	-
	\frac{1}{4}\left(\constthreee \beta^2 + \constthreef \delta^2 + \constthreekin\beta\delta \right)
	\,G_{AB} \,\mathcal{E}^{ABDE}\,G_{DE}
	\\
	&
	-
	\frac{1}{4}\left(2\constthreee \alpha\,\beta + 2\constthreef \gamma\,\delta + \constthreekin(\alpha\delta+\beta\gamma)\right)
	\,M_{AB} \,\mathcal{E}^{ABDE}\,G_{DE},
	\end{split}
	\end{align}
	where $\supset$ denotes that some terms in the Lagrangian are omitted, and imposing the above condition, as well as
	\begin{align}
	\begin{cases}
	\constthreee \alpha^2 + \constthreef \gamma^2 + \constthreekin\alpha\gamma = -1\\
	\constthreee \beta^2 + \constthreef \delta^2 + \constthreekin\beta\delta = -1\\
	2\constthreee \alpha\,\beta + 2\constthreef \gamma\,\delta + \constthreekin(\alpha\delta+\beta\gamma) = 0
	\end{cases}
	\end{align}
	results in four possible solutions for the involved parameters which differ only by the verse associated to the selected linear combinations in field space. In particular, the choice
	\begin{align}
	\label{eq:goodsub}
	\begin{cases}
	\alpha = -\dfrac{\constthreekin m_\ee + 2\constthreef m_\ff}{\sqrt{\left(\constthreekin^2 -4\constthreee \constthreef \right)\left(\constthreee m_\ee^2 + \constthreef m_\ff^2 + \constthreekin m_\ee m_\ff  \right)} }
	\\
	\beta = \dfrac{ m_\ee}{\sqrt{-\left(\constthreee m_\ee^2 + \constthreef m_\ff^2 +  m_\ee m_\ff  \right)} }\\
	\gamma = \dfrac{\constthreekin m_\ff + 2\constthreee m_\ee}{\sqrt{\left(\constthreekin^2 -4\constthreee \constthreef \right)\left(\constthreee m_\ee^2 + \constthreef m_\ff^2 + \constthreekin m_\ee m_\ff  \right)} }\\
	\delta = \dfrac{ m_\ff}{\sqrt{-\left(\constthreee m_\ee^2 + \constthreef m_\ff^2 + \constthreekin m_\ee m_\ff  \right)} }
	\end{cases}
	\end{align}
	correctly reduces to the (inverse of the) rotation matrix in eq.$\eqref{eq:eigrot}$. Real and finite coefficients are obtained when
	\begin{align}
	\label{eq:k3cond}
	\begin{cases}
	\constthreee m_\ee^2 + \constthreef m_\ff^2 + \constthreekin m_\ee m_\ff<0 \\
	\constthreekin^2 -4\constthreee \constthreef <0
	\end{cases}
	\end{align}
	which also ensure that the transformation is invertible:
	\begin{align}
	\det
	\begin{pmatrix}
	\alpha & \beta \\ \gamma &\delta
	\end{pmatrix}
	\overset{\text{eq.}\eqref{eq:goodsub}}{=}
	\frac{2\sqrt{-\left(\constthreee m_\ee^2 + \constthreef m_\ff^2 + \constthreekin m_\ee m_\ff  \right)}}{\sqrt{\left(\constthreekin^2 -4\constthreee \constthreef \right)\left(\constthreee m_\ee^2 + \constthreef m_\ff^2 + \constthreekin m_\ee m_\ff  \right)}}
	\end{align}
	We notice the following particular values of $\constthreekin$:
	\begin{itemize}
		\item $\constthreekin = \pm 2 \sqrt{\constthreee \constthreef}$. In this case the kinetic term acquires the form \\ \noindent $\LG \supset A^2  \,M_{AB} \,\tilde{\mathcal{E}}^{ABDE}\,M_{DE} + B^2 \,G_{AB} \,\tilde{\mathcal{E}}^{ABDE}\,G_{DE} + 2AB \,G_{AB} \,\tilde{\mathcal{E}}^{ABDE}\,M_{DE}$, implying that a vanishing mixing term necessarily requires at least either of the kinetic terms to vanish.
		\\
		\item  $\constthreekin = - (\constthreee m_\ee^2 + \constthreef m_\ff^2)/ m_e m_f$.  Once eq.$\eqref{eq:noGmass}$ is imposed to prevent a mass term for the $G_{AB}$ perturbation, this particular value of $\constthreekin$ nullifies the kinetic term of the same perturbation.
	\end{itemize}

	Excluding the above values of $\constthreekin$, the Lagrangian of the symmetric sector simply becomes
	\begin{align}
	\LG^{hh} =
	\frac{1}{4}
	\,M_{AB} \,\mathcal{E}^{ABDE}\,M_{DE}
	+
	\frac{1}{4} \,G_{AB} \,\mathcal{E}^{ABDE}\,G_{DE}-
	\frac{m_{FP}^2}{4}
	\left(M_{AB}M^{AB} - M^2 \right)
	\end{align}
	where the involved perturbations are determined from the original $\he_{AB}$ and $\hf_{AB}$ fields through the inverse of eq.$\eqref{eq:gen_t}$ and where we have defined
	\begin{align}
	m_{FP}^2:=m_{int}^4\left(\frac{\alpha}{m_\ee}-\frac{\gamma}{m_\ff}\right)^2=m_{int}^4\frac{4 \left(\constthreee m_\ee^2+ \constthreekin m_\ee m_\ff + \constthreef m_\ff^2\right)}{m_\ee^2 m_\ff^2 \left(\constthreekin^2-4\constthreee \constthreef\right)}\,.
	\end{align}
	Notice that the Fierz-Pauli mass is positive defined if the conditions in eq.$\eqref{eq:k3cond}$ are satisfied, and becomes vanishing or singular in correspondence of the two values of $\constthreekin$ that make the transformation singular. This seems to imply that if a fluctuation does not propagate, the other must necessarily be massless.

	\subsubsection{Antisymmetric sector}

	Analogously, for the antisymmetric perturbations we have
	\begin{align}
	\LG^{bb}
	&=
	-\frac14\,\be_{AB} \, (2\consttwoe+\constthreee) \,\mathcal{E}^{ABDE}\,\be_{DE}
	-
	\frac14\,\bff_{AB} \,(2\consttwof+\constthreef) \,\mathcal{E}^{ABDE}\,\bff_{DE}
	-
	\frac14\,\be_{AB} \,(2\consttwokin+\constthreekin) \,\mathcal{E}^{ABDE}\,\bff_{DE}
	\\ \nn
	&
	+\frac{m_{int}^4}{4}
	\left(\frac{\be_{AB}}{m_\ee}-\frac{\bff_{AB}}{m_\ff}\right)\left(\frac{\be^{AB}}{m_\ee}-\frac{\bff^{AB}}{m_\ff}\right)
	,
	\end{align}
	where we used $\left(3\beta_1 + 4\beta_2 +3\beta_3 \right)=24$. In order to diagonalize the above Lagrangian we introduce the states $E_{AB}$ and $F_{AB}$ such that
	\begin{align}
	\begin{pmatrix}
	\be_{AB}\\ \bff_{AB}
	\end{pmatrix}
	=:
	\begin{pmatrix}
	a &b \\ c&d
	\end{pmatrix}
	\begin{pmatrix}
	E_{AB}\\F_{AB}
	\end{pmatrix}\,,
	\end{align}
	and express the mass term as
	\begin{align}
	\LG^{bb}\supset\frac{m^4_{int}}{4}
	\left[\left(\frac{a}{m_\ee}-\frac{c}{m_\ff}\right)E_{AB} + \left(\frac{b}{m_\ee}-\frac{d}{m_\ff}\right)F_{AB}\right]\left[\left(\frac{a}{m_\ee}-\frac{c}{m_\ff}\right)E^{AB} + \left(\frac{b}{m_\ee}-\frac{d}{m_\ff}\right)F^{AB}\right]\,.
	\end{align}
	Imposing the absence of mass mixing then forces, again, the presence of a massless state. In analogy to the symmetric case, we choose here
	\begin{align}
	\frac{b}{m_\ee}\req\frac{d}{m_\ff}
	\end{align}
	so that
	\begin{align}
	\LG^{bb}\supset\frac{m^4_{int}}{4}\left(\frac{a}{m_\ee}-\frac{c}{m_\ff}\right)^2
	E_{AB}E^{AB}=: \frac{\mu_{FP}^2}{4}E_{AB}E^{AB}\,.
	\end{align}
	Notice that $\mu^2_{FP}\geq0$ and consequently the mass term appears in the Lagrangian inevitably with the wrong sign. Leaving this issue aside for the moment, we proceed to express the kinetic term in terms of the newly defined perturbations
	\begin{align}
	\LG^{bb}&\supset
	-\frac{1}{4}\left((2\consttwoe+\constthreee) \, a^2 + (2\consttwof+\constthreef) \, c^2 + (2\consttwokin+\constthreekin)\,ac \right)
	\,E_{AB} \,{\mathcal{E}}^{ABDE}\,E_{DE}
	\\
	&-
	\frac{1}{4}\left((2\consttwoe+\constthreee) \, b^2 + (2\consttwof+\constthreef) \, d^2 + (2\consttwokin+\constthreekin)\,bd \right)
	\,F_{AB} \,{\mathcal{E}}^{ABDE}\,F_{DE}
	+ \nn\\
	&
	-\frac{1}{4}\left(2(2\consttwoe+\constthreee) \, ab + 2(2\consttwof+\constthreef)  \, cd +(2\consttwokin+\constthreekin)\,(ad+bc)\right)
	\,E_{AB} \,{\mathcal{E}}^{ABDE}\,F_{DE}
	\end{align}\,.
	Requiring now
	\begin{align}
	\begin{cases}
	(2\consttwoe+\constthreee) \, a^2 + (2\consttwof+\constthreef) \, c^2 + (2\consttwokin+\constthreekin)\,ac  =-1
	\\
	(2\consttwoe+\constthreee) \, b^2 + (2\consttwof+\constthreef) \, d^2 + (2\consttwokin+\constthreekin)\,bd = -1
	\\
	2(2\consttwoe+\constthreee) \, ab + 2(2\consttwof+\constthreef)  \, cd +(2\consttwokin+\constthreekin)\,(ad+bc) = 0
	\end{cases}
	\end{align}
	results in
	\begin{align}
	\begin{cases}
	a = -\dfrac{(2\consttwokin+\constthreekin) m_\ee + 2(2\consttwof+\constthreef) m_\ff}{\sqrt{\left((2\consttwokin+\constthreekin)^2 -4(2\consttwoe+\constthreee) (2\consttwof+\constthreef) \right)\left((2\consttwoe+\constthreee) m_\ee^2 +(2\consttwof+\constthreef) m_\ff^2 + (2\consttwokin+\constthreekin) m_\ee m_\ff  \right)} }
	\\
	b = \dfrac{ m_\ee}{\sqrt{-\left((2\consttwoe+\constthreee) m_\ee^2 +(2\consttwof+\constthreef) m_\ff^2 + (2\consttwokin+\constthreekin) m_\ee m_\ff   \right)} }\\
	c = \dfrac{(2\consttwokin+\constthreekin) m_\ff + 2(2\consttwoe+\constthreee)  m_\ee}{\sqrt{\left((2\consttwokin+\constthreekin)^2 -4(2\consttwoe+\constthreee) (2\consttwof+\constthreef) \right)\left((2\consttwoe+\constthreee) m_\ee^2 +(2\consttwof+\constthreef) m_\ff^2 + (2\consttwokin+\constthreekin) m_\ee m_\ff  \right)} }\\
	d = \dfrac{ m_\ff}{\sqrt{-\left((2\consttwoe+\constthreee) m_\ee^2 +(2\consttwof+\constthreef) m_\ff^2 + (2\consttwokin+\constthreekin) m_\ee m_\ff   \right)} }
	\end{cases}
	\end{align}
	and every other result can, likewise, be derived from those of the symmetric sector via the substitution $\constthreekin \to 2\consttwokin + \constthreekin$, $\constthreee \to 2\consttwoe + \constthreee$, $\constthreef \to 2\consttwof+\constthreef$.

	As in the case with vanishing kinetic mixing, the general theory seems to admit the propagation of a ghost-like massive antisymmetric perturbation. Although it is certainly possible to force the kinetic terms of this problematic perturbation to vanish through a proper choice of the coefficients, we argue below that this might not be necessary. As antisymmetric perturbations correspond, in essence, to the propagation of scalar fields, the presence of a negative mass term could simply indicate the emergence of a spontaneously broken symmetry. In fact, for the sake of the viability of the theory, it could be enough that the terms of higher order in the perturbative expansion give rise to self interactions of the scalar field which result in a potential bounded from below. The situation is analogous to that of the Higgs boson in the standard model, where the negative mass term of the field triggers the electroweak symmetry breaking.

	\section{Nonlinear analysis of teleparallel bigravity}
	\label{sec:Nonlinear}
	We saw in the previous section that some teleparallel bigravity theories contain ghosts. Still, it is of interest to understand their symmetries and view them as toy models. For instance, the terms with derivative interactions might be constructed in a viable way if we allow for a construction with complex tetrads similar to the theory considered in \cite{Apolo:2016ort,Apolo:2016vkn}. We dedicate this section to investigate teleparallel bigravity at the nonlinear level. For this we make a 3+1 decomposition and find primary constraints for the action considered in the linear analysis.

	\subsection{Primary constraints}
	\label{sec:PrimaryConstraints}
	We remind ourselves that the linear analysis applies to the action \eqref{GenBiGrav} with the supermetrics defined by \eqref{eq:supmetee},\eqref{eq:supmetkin}, and \eqref{eq:supmetff}. Note that the interaction potential is independent of velocities in the fields and, hence, do not contribute to the conjugate momenta. Furthermore, $\overset{\ee}\LG$ and $\overset{\ff}\LG$ are just new general relativity terms which are already known from the literature \cite{Blixt:2018znp,Blixt:2019ene,Blixt:2020ekl,Blagojevic:2000qs,Mitric:2019rop}. Hence, we can determine the primary constraints by making the 3+1 decomposition of $\overset{kin}\LG$ and calculate its primary constraints and add this information. The lapse and shift are denoted by $\lapseh, \ \shifth^i$, respectively and the induced metric is denoted by $\induceh$, whereas the normal vector components to the hypersurface of constant time slices is denoted by $\normhh^A$. The Lagrangian 3+1 decomposition of $\overset{kin}\LG$ is
	\begin{align}\label{eq:Lkin}
	\begin{split}
	\overset{kin}\LG&=\frac{m_{ef}^2\sqrt{\induceh}}{\lapseh}\accentset{\ee}{T}^A{}_{i0}\accentset{\ff}{T}^B{}_{j0}\Mkin^{i\ j}_{\ A\ B}
	\\&+\frac{m_{ef}^2\sqrt{\induceh}}{\lapseh}\left(\accentset{\ee}{T}^A{}_{i0}\accentset{\ff}{T}^B{}_{kl}+\accentset{\ff}{T}^A{}_{i0}\accentset{\ee}{T}^B{}_{kl}\right)\left[\Mkin^{i \ l}_{\ A \ B}\shifth^{k}-\lapseh\induceh^{il}\left(\consttwokin\normhh_{B}\cotetradmean_A{}^k+ \constthreekin\normhh_{A}\cotetradmean_B{}^k \right)\right]
	\\&+ \frac{m_{ef}^2\sqrt{\induceh}}{\lapseh}\left(\accentset{\ee}{T}^A{}_{ij}\accentset{\ff}{T}^B{}_{kl}+\accentset{\ff}{T}^A{}_{ij}\accentset{\ee}{T}^B{}_{kl}\right)\left[\frac{1}{2}\Mkin^{\ i\ k}_{\ A\ B}\shifth^{j}\shifth^{l}- \lapseh \induceh^{jl}\normhh_A \left(\consttwokin \cotetradmean_B{}^i\shifth^{k}+ \constthreekin\cotetradmean_B{}^k\shifth^{i} \right)\right]
	\\&+\lapseh m_{ef}^2\sqrt{\induceh}\cdot \spatialhh,
	\end{split}
	\end{align}
	where
	\begin{align}
	\begin{split}
	\Mkin^{i \ j}_{ \ A \ B}&=4\lapseh^{2}\accentset{kin}{G}_{AB}^{\ \ \ \  i0j0}
	\\&=2\constonekin\induceh^{ij}\eta_{AB}-(\consttwokin+\constthreekin)\normhh_{A}\normhh_{B}\induceh^{ij}+\consttwokin\cotetradmean_A{}^j\cotetradmean_B{}^i+\constthreekin\cotetradmean_A{}^i\cotetradmean_B{}^j,
	\end{split}
	\end{align}
	and
	\begin{align}
	\begin{split}
	\spatialhh\equiv -\constonekin\eta_{AB}\accentset{\ee}{T}^A{}_{ij}\accentset{\ff}{T}^B{}_{kl}\induceh^{ik}\induceh^{jl}-\consttwokin\cotetradmean_A{}^i \cotetradmean_B{}^j \accentset{\ee}{T}^A{}_{kj}\accentset{\ff}{T}^B{}_{li}\induceh^{kl}-\constthreekin\cotetradmean_A{}^i\cotetradmean_B{}^j \induceh^{kl}\accentset{\ee}{T}^A{}_{ki}\accentset{\ff}{T}^B{}_{lj}.
	\end{split}
	\end{align}
	The conjugate momenta hence becomes
	\begin{align}
	\frac{\conje_A{}^i}{m_{ef}^2}=\frac{\sqrt{\induceh}}{\lapseh}\dot{f}^B{}_j\Mkin^{i \ j}_{\ A \ B}-\frac{\sqrt{\induceh}}{\lapseh}\covf_jf^B{}_0\Mkin^{i \ j}_{\ A \ B}-\frac{\sqrt{\induceh}}{\lapseh}\accentset{\ff}{T}^B{}_{kl}\left[\Mkin^{i \ l}_{\ A \ B}\shifth^k-\lapseh\induceh^{il}\left(\consttwokin\normhh_B\cotetradmean_A{}^k+\constthreekin\normhh_A\cotetradmean_B{}^k\right)\right],
	\\ 	\frac{\conjf_A{}^i}{m_{ef}^2}=\frac{\sqrt{\induceh}}{\lapseh}\dot{e}^B{}_j\Mkin^{i \ j}_{\ A \ B}-\frac{\sqrt{\induceh}}{\lapseh}\covg_je^B{}_0\Mkin^{i \ j}_{\ A \ B}-\frac{\sqrt{\induceh}}{\lapseh}\accentset{\ee}{T}^B{}_{kl}\left[\Mkin^{i \ l}_{\ A \ B}\shifth^k-\lapseh\induceh^{il}\left(\consttwokin\normhh_B\cotetradmean_A{}^k+\constthreekin\normhh_A\cotetradmean_B{}^k\right)\right].
	\end{align}
	We decompose everything into irreducible parts, with respect to $\cotetradmean$, as done in \cite{Blagojevic:2000qs,Blixt:2018znp}.
	\begin{align} \label{eq:irrE}
	\dot{\ee}^A{}_i&={}^{\mathcal{V}}\dot{\ee}_i \normhh^A+{}^{\mathcal{A}}\dot{\ee}_{ji}\induceh^{kj}\cotetradmean^A{}_k+{}^{\mathcal{S}}\dot{\ee}_{ji}\induceh^{kj}\cotetradmean^A{}_k+{}^{\mathcal{T}}\dot{\ee}\cotetradmean^A{}_i\\ \label{eq:irrF}
	\dot{\ff}^A{}_i&={}^{\mathcal{V}}\dot{\ff}_i \normhh^A+{}^{\mathcal{A}}\dot{\ff}_{ji}\induceh^{kj}\cotetradmean^A{}_k+{}^{\mathcal{S}}\dot{\ff}_{ji}\induceh^{kj}\cotetradmean^A{}_k+{}^{\mathcal{T}}\dot{\ff}\cotetradmean^A{}_i\\
	\label{eq:irrconjE}
	\conje^A{}_i&={}^{\mathcal{V}}\conje_i \normhh^A+{}^{\mathcal{A}}\conje_{ji}\induceh^{kj}\cotetradmean^A{}_k+{}^{\mathcal{S}}\conje_{ji}\induceh^{kj}\cotetradmean^A{}_k+{}^{\mathcal{T}}\conje \cotetradmean^A{}_i\\
	\label{eq:irrconjF}
	\conjf^A{}_i&={}^{\mathcal{V}}\conjf_i \normhh^A+{}^{\mathcal{A}}\conjf_{ji}\induceh^{kj}\cotetradmean^A{}_k+{}^{\mathcal{S}}\conjf_{ji}\induceh^{kj}\cotetradmean^A{}_k+{}^{\mathcal{T}}\conjf \cotetradmean^A{}_i.
	\end{align}
	The inversion for a general theory \eqref{GenBiGrav} is quite involved and lies beyond the scope of this article. However, we can, as a toy example, derive the inversion formula between the velocities and conjugate momenta for the subtheory $\constonee=\consttwoe=\constthreee=\constonef=\consttwof=\constthreef=0$, for which the results are very similar to \cite{Blixt:2018znp}. First we define the sources which are independent of velocities:
	\begin{align}
	\label{gSource}
	\sourceg_A{}^i=	\lapseh\frac{\conje_A{}^i}{m_{ef}^2\sqrt{\induceh}}+\covf_j\ff^B{}_0\Mkin^{i \ j}_{\ A \ B}+\accentset{\ff}{T}^B{}_{kl}\left[\Mkin^{i \ l}_{\ A \ B}\shifth^k-\lapseh \consttwokin\induceh^{il}\normhh_B\cotetradmean_A{}^k-\lapseh\constthreekin\induceh^{il}\normhh_A\cotetradmean_B{}^k\right]=\dot{\ff}^B{}_j\Mkin^{i \ j}_{\ A \ B},
	\end{align}
	\begin{align}
	\label{fSource}
	\sourcef_A{}^i=\lapseh\frac{\conjf_A{}^i}{m_{ef}^2\sqrt{\induceh}}+\covg_j\ee^B{}_0\Mkin^{i \ j}_{\ A \ B}+\accentset{\ee}{T}^B{}_{kl}\left[\Mkin^{i \ l}_{\ A \ B}\shifth^k-\lapseh\consttwokin\induceh^{il}\normhh_B\cotetradmean_A{}^k-\lapseh\constthreekin\induceh^{il}\normhh_A\cotetradmean_B{}^k\right]=\dot{\ee}^B{}_j\Mkin^{i \ j}_{\ A \ B},
	\end{align}
	where $\Mkin^{i \ j}_{\ A \ B}$ and its inverse can be written in irreducible parts:
	\begin{align}
	\Mkin^{i \ j}_{\ A \ B}={}^{\mathcal{V}}\Mkin^{i j}\normhh_A \normhh_B+{}^{\mathcal{A}}\Mkin^{[i r][js]} \cotetradmean^C{}_r\eta_{AC}\cotetradmean^D{}_s\eta_{BD}+{}^{\mathcal{S}}\Mkin^{[i r][js]} \cotetradmean^C{}_r\eta_{AC}\cotetradmean^D{}_s\eta_{BD} +{}^{\mathcal{T}}\Mkin\cotetradmean_A{}^i\cotetradmean_B{}^j,
	\end{align}
	which yields:
	\begin{align}\label{MIrr}
	\begin{split}
	\Mkin^{i \ j}_{\ A \ B}&=-2\kinveccoeffimposeconstr \normhh_A\normhh_B\induceh^{ij}+2\kinantcoeffimposeconstr \induceh^{i[j}\induceh^{s]r}\cotetradmean^C{}_r\eta_{AC}\cotetradmean^D{}_s\eta_{BD}+2\kinsymcoeffimposeconstr \left(\induceh^{i(j}\induceh^{s)r}+\frac{1}{3}\induceh^{ir}\induceh^{js}\right)\cotetradmean^C{}_r\eta_{AC}\cotetradmean^D{}_s\eta_{BD}\\
	&+\frac{2}{3}\kintrcoeffimposeconstr \cotetradmean_A{}^i\cotetradmean_B{}^j,
	\end{split}
	\end{align}
	where
	\begin{align}
	\kinveccoeffimposeconstr&=2\constonekin+\consttwokin+\constthreekin,\\
	\kinantcoeffimposeconstr&=2\constonekin-\consttwokin\\
	\kinsymcoeffimposeconstr&=2\constonekin+\consttwokin\\
	\kintrcoeffimposeconstr&=2\constonekin+\consttwokin+3\constthreekin.
	\end{align}
	Putting $\kincoeffimposeconstr=0 \ (\mathcal{I}\in \{\mathcal{V},\mathcal{A},\mathcal{S},\mathcal{T}\}$) we get the following primary constraints ${}^{\mathcal{I}}C \approx 0$ and ${}^{\mathcal{I}}\tilde{C} \approx 0$:
	\begin{align}
	\begin{split}
	\CkineVec^i=\frac{{}^{\mathcal{V}}\conje^i}{\sqrt{\induceh}}-\constthreekin m_{ef}^2\accentset{\ff}{T}^B{}_{kl}\induceh^{il}\cotetradmean_B{}^k \approx 0,\\
	\CkinfVec^i=\frac{{}^{\mathcal{V}}\conjf^i}{{\sqrt{\induceh}}}-\constthreekin m_{ef}^2\accentset{\ee}{T}^B{}_{kl}\induceh^{il}\cotetradmean_B{}^k \approx 0.
	\end{split}
	\end{align}
	\begin{align}
	\begin{split}
	\CkineAn_{ij}=\frac{{}^{\mathcal{A}}\conje_{ij}}{\induceh}-\consttwokin m_{ef}^2\accentset{f}{T}^B{}_{ij}\normhh_B \approx 0,\\
	\CkinfAn_{ij}=\frac{{}^{\mathcal{A}}\conjf_{ij}}{\sqrt{\induceh}}-\consttwokin m_{ef}^2\accentset{\ee}{T}^B{}_{ij}\normhh_B \approx 0.
	\end{split}
	\end{align}
	\begin{align}
	\begin{split}
	\CkineSym_{ij}=\frac{{}^{\mathcal{S}}\conje_{ij}}{\sqrt{\induceh}}\approx 0,\\
	\CkinfSym_{ij}=\frac{{}^{\mathcal{S}}\conjf_{ij}}{\sqrt{\induceh}}\approx 0.
	\end{split}
	\end{align}
	\begin{align}
	\begin{split}
	\CkineTr=\frac{{}^{\mathcal{T}}\conje}{\sqrt{\induceh}} \approx 0,\\
	\CkinfTr=\frac{{}^{\mathcal{T}}\conjf}{\sqrt{\induceh}} \approx 0.
	\end{split}
	\end{align}
	We now want to include the contribution from $\overset{\ee}\LG$ and $\overset{\ff}\LG$. The 3+1 decomposition of this Lagrangian reads
	\begin{align}\label{eq:Lkin+2copies}
	\begin{split}
	\overset{\ee}{\LG}+\overset{kin}\LG+\overset{\ff}{\LG}&=\frac{m_e^2\sqrt{\inducee}}{\lapsee}\accentset{\ee}{T}^A{}_{i0}\accentset{\ee}{T}^B{}_{j0}\accentset{\ee}{M}^{i\ j}_{\ A \ B}+\frac{m_{ef}^2\sqrt{\induceh}}{\lapseh}\accentset{\ee}{T}^A{}_{i0}\accentset{\ff}{T}^B{}_{j0}\Mkin^{i\ j}_{\ A\ B}+\frac{m_f^2\sqrt{\inducef}}{\lapsef}\accentset{\ff}{T}^A{}_{i0}\accentset{\ff}{T}^B{}_{j0}\accentset{\ff}{M}^{i\ j}_{\ A \ B}
	\\&+\frac{m_e^2\sqrt{\inducee}}{\lapsee}\accentset{\ee}{T}^A{}_{i0}\accentset{\ee}{T}^B{}_{kl}\left[\accentset{\ee}{M}^{i\ l}_{ \ A \ B}\shifte^k-2\lapsee \inducee^{il}\left(\consttwoe \normee_{B}\ee_A{}^k+\constthreee\normee_{A}\ee_B{}^k\right)\right]
	\\&+\frac{m_{ef}^2\sqrt{\induceh}}{\lapseh}\left(\accentset{\ee}{T}^A{}_{i0}\accentset{\ff}{T}^B{}_{kl}+\accentset{\ff}{T}^A{}_{i0}\accentset{\ee}{T}^B{}_{kl}\right)\left[\Mkin^{i \ l}_{\ A \ B}\shifth^{k}-\lapseh\induceh^{il}\left(\consttwokin\normhh_{B}\cotetradmean_A{}^k+ \constthreekin\normhh_{A}\cotetradmean_B{}^k \right)\right]
	\\&+\frac{m_f^2\sqrt{\inducef}}{\lapsef}\accentset{\ff}{T}^A{}_{i0}\accentset{\ff}{T}^B{}_{kl}\left[\accentset{\ff}{M}^{i\ l}_{ \ A \ B}\shiftf^k-2\lapsef \inducef^{il}\left(\consttwof \normff_{B}\ff_A{}^k+\constthreef\normff_{A}\ff_B{}^k\right)\right]
	\\&+\frac{m_e^2\sqrt{\inducee}}{\lapsee}\accentset{\ee}{T}^A{}_{ij}\accentset{\ee}{T}^B{}_{kl}\shifte^i\left[\frac{1}{2}\accentset{\ee}{M}^{j \ l}_{\ A \ B}\shifte^k-2\lapsee \inducee^{jl}\left(\consttwoe\normee_{B}\ee_A{}^k+\constthreee\normee_{A}\ee_B{}^k\right)\right]
	\\&+ \frac{m_{ef}^2\sqrt{\induceh}}{\lapseh}\left(\accentset{\ee}{T}^A{}_{ij}\accentset{\ff}{T}^B{}_{kl}+\accentset{\ff}{T}^A{}_{ij}\accentset{\ee}{T}^B{}_{kl}\right)\left[\frac{1}{2}\Mkin^{\ i\ k}_{\ A\ B}\shifth^{j}\shifth^{l}- \lapseh \induceh^{jl}\normhh_A \left(\consttwokin \cotetradmean_B{}^i\shifth^{k}+ \constthreekin\cotetradmean_B{}^k\shifth^{i} \right)\right]
	\\&+\frac{m_f^2\sqrt{\inducef}}{\lapsef}\accentset{\ff}{T}^A{}_{ij}\accentset{\ff}{T}^B{}_{kl}\shiftf^i\left[\frac{1}{2}\accentset{\ff}{M}^{j \ l}_{\ A \ B}\shiftf^k-2\lapsef \inducef^{jl}\left(\consttwof\normff_{B}\ff_A{}^k+\constthreef\normff_{A}\ff_B{}^k\right)\right]
	\\&+\lapsee m_e^2\sqrt{\inducee}\cdot \spatialee+\lapseh m_{ef}^2 \sqrt{\induceh}\cdot \spatialhh+\lapsef m_f^2\sqrt{\inducef}\cdot \spatialff,
	\end{split}
	\end{align}
	where we introduce
	\begin{align}
	\begin{split}
	\accentset{\ee}{M}^{i \ j}_{ \ A \ B}&=4\lapsee^{2}\accentset{\ee}{G}_{AB}^{\ \ \ \  i0j0}
	\\&=2\constonee\inducee^{ij}\eta_{AB}-(\consttwoe+\constthreee)\normee_{A}\normee_{B}\inducee^{ij}+\consttwoe \ee_A{}^j \ee_B{}^i+\constthreee \ee_A{}^i \ee_B{}^j,
	\end{split}
	\end{align}
	\begin{align}
	\begin{split}
	\accentset{\ff}{M}^{i \ j}_{ \ A \ B}&=4\lapsef^{2}\accentset{\ff}{G}_{AB}^{\ \ \ \  i0j0}
	\\&=2\constonef\inducef^{ij}\eta_{AB}-(\consttwof+\constthreef)\normff_{A}\normff_{B}\inducef^{ij}+\consttwof \ff_A{}^j \ff_B{}^i+\constthreef \ff_A{}^i \ff_B{}^j,
	\end{split}
	\end{align}
	\begin{align}
	\begin{split}
	\spatialee\equiv -\constonee\eta_{AB}\accentset{\ee}{T}^A{}_{ij}\accentset{\ee}{T}^B{}_{kl}\inducee^{ik}\inducee^{jl}-\consttwoe \ee_A{}^i \ee_B{}^j \accentset{\ee}{T}^A{}_{kj}\accentset{\ee}{T}^B{}_{li}\inducee^{kl}-\constthreee \ee_A{}^i \ee_B{}^j \inducee^{kl}\accentset{\ee}{T}^A{}_{ki}\accentset{\ee}{T}^B{}_{lj},
	\end{split}
	\end{align}
	\begin{align}
	\begin{split}
	\spatialff\equiv -\constonef\eta_{AB}\accentset{\ff}{T}^A{}_{ij}\accentset{\ff}{T}^B{}_{kl}\inducef^{ik}\inducef^{jl}-\consttwof \ff_A{}^i \ff_B{}^j \accentset{\ff}{T}^A{}_{kj}\accentset{\ff}{T}^B{}_{li}\inducef^{kl}-\constthreef \ff_A{}^i \ff_B{}^j \inducef^{kl}\accentset{\ff}{T}^A{}_{ki}\accentset{\ff}{T}^B{}_{lj}.
	\end{split}
	\end{align}
	The conjugate momenta for this action are
	\begin{align}\begin{split}
	\frac{\conje_A{}^i}{m_{ef}^2}&=\frac{\sqrt{\induceh}}{\lapseh}\dot{\ff}^B{}_j\Mkin^{i \ j}_{\ A \ B}-\frac{\sqrt{\induceh}}{\lapseh}\covf_j\ff^B{}_0\Mkin^{i \ j}_{\ A \ B}-\frac{\sqrt{\induceh}}{\lapseh}\accentset{\ff}{T}^B{}_{kl}\left[\Mkin^{i \ l}_{\ A \ B}\shifth^k-\lapseh\induceh^{il}\left(\consttwokin\normhh_B\cotetradmean_A{}^k+\constthreekin\normhh_A\cotetradmean_B{}^k\right)\right]
	\\ &+\frac{\sqrt{\inducee}}{\lapsee}\dot{\ee}^B{}_j\accentset{\ee}{M}^{i \ j}_{\ A \ B}-\frac{\sqrt{\inducee}}{\lapsee}\covg_j\ee^B{}_0\accentset{\ee}{M}^{i \ j}_{\ A \ B}-\frac{\sqrt{\inducee}}{\lapsee}\accentset{\ee}{T}^B{}_{kl}\left[\accentset{\ee}{M}^{i \ l}_{\ A \ B}\shifte^k-\lapsee\inducee^{il}\left(\consttwoe\normee_B \ee_A{}^k+\constthreee\normee_A\ee_B{}^k\right)\right],
	\end{split}
	\end{align}
	\begin{align}
	\begin{split}	\frac{\conjf_A{}^i}{m_{ef}^2}&=\frac{\sqrt{\induceh}}{\lapseh}\dot{\ee}^B{}_j\Mkin^{i \ j}_{\ A \ B}-\frac{\sqrt{\induceh}}{\lapseh}\covg_j\ee^B{}_0\Mkin^{i \ j}_{\ A \ B}-\frac{\sqrt{\induceh}}{\lapseh}\accentset{\ee}{T}^B{}_{kl}\left[\Mkin^{i \ l}_{\ A \ B}\shifth^k-\lapseh\induceh^{il}\left(\consttwokin\normhh_B\cotetradmean_A{}^k+\constthreekin\normhh_A\cotetradmean_B{}^k\right)\right]\\
	&+\frac{\sqrt{\inducef}}{\lapsef}\dot{\ff}^B{}_j\accentset{\ff}{M}^{i \ j}_{\ A \ B}-\frac{\sqrt{\inducef}}{\lapsef}\covf_j\ff^B{}_0\accentset{\ff}{M}^{i \ j}_{\ A \ B}-\frac{\sqrt{\inducef}}{\lapsef}\accentset{\ff}{T}^B{}_{kl}\left[\accentset{\ff}{M}^{i \ l}_{\ A \ B}\shiftf^k-\lapsef\inducef^{il}\left(\consttwof\normff_B \ff_A{}^k+\constthreef\normff_A\ff_B{}^k\right)\right].
	\end{split}
	\end{align}
	We can, as in the $\overset{kin}{\LG}$ case, define sources which are independent of velocities so that
	\begin{align}\label{eq:SourcegTot}
	\accentset{\ee}{\tilde{S}}_A{}^i\equiv \frac{\sqrt{\induceh}}{\lapseh}\dot{\ff}^B{}_j\Mkin^{i \ j}_{\ A\ B}+\frac{\sqrt{\inducee}}{\lapsee}\dot{\ee}^B{}_j\accentset{\ee}{M}^{i\ j}_{\ A \ B},\\\label{eq:SourcefTot}
	\accentset{\ff}{\tilde{S}}_A{}^i\equiv \frac{\sqrt{\induceh}}{\lapseh}\dot{\ee}^B{}_j\Mkin^{i \ j}_{\ A\ B}+\frac{\sqrt{\inducef}}{\lapsef}\dot{\ff}^B{}_j\accentset{\ff}{M}^{i\ j}_{\ A \ B},
	\end{align}
	with
	\begin{align}\begin{split} \label{eq:invertvel}
	\accentset{\ee}{\tilde{S}}_A{}^i&=\frac{\conje_A{}^i}{m_{ef}^2}+\frac{\sqrt{\induceh}}{\lapseh}\covf_j\ff^B{}_0\Mkin^{i \ j}_{\ A \ B}+\frac{\sqrt{\induceh}}{\lapseh}\accentset{\ff}{T}^B{}_{kl}\left[\Mkin^{i \ l}_{\ A \ B}\shifth^k-\lapseh\induceh^{il}\left(\consttwokin\normhh_B\cotetradmean_A{}^k+\constthreekin\normhh_A\cotetradmean_B{}^k\right)\right]
	\\ &+\frac{\sqrt{\inducee}}{\lapsee}\covg_j\ee^B{}_0\accentset{\ee}{M}^{i \ j}_{\ A \ B}+\frac{\sqrt{\inducee}}{\lapsee}\accentset{\ee}{T}^B{}_{kl}\left[\accentset{\ee}{M}^{i \ l}_{\ A \ B}\shifte^k-\lapsee\inducee^{il}\left(\consttwoe\normee_B \ee_A{}^k+\constthreee\normee_A\ee_B{}^k\right)\right],\\
	\accentset{\ff}{\tilde{S}}_A{}^i&=\frac{\conjf_A{}^i}{m_{ef}^2}+\frac{\sqrt{\induceh}}{\lapseh}\covg_j\ee^B{}_0\Mkin^{i \ j}_{\ A \ B}+\frac{\sqrt{\induceh}}{\lapseh}\accentset{\ee}{T}^B{}_{kl}\left[\Mkin^{i \ l}_{\ A \ B}\shifth^k-\lapseh\induceh^{il}\left(\consttwokin\normhh_B\cotetradmean_A{}^k+\constthreekin\normhh_A\cotetradmean_B{}^k\right)\right]\\
	&+\frac{\sqrt{\inducef}}{\lapsef}\covf_j\ff^B{}_0\accentset{\ff}{M}^{i \ j}_{\ A \ B}+\frac{\sqrt{\inducef}}{\lapsef}\accentset{\ff}{T}^B{}_{kl}\left[\accentset{\ff}{M}^{i \ l}_{\ A \ B}\shiftf^k-\lapsef\inducef^{il}\left(\consttwof\normff_B \ff_A{}^k+\constthreef\normff_A\ff_B{}^k\right)\right],
	\end{split}
	\end{align}
	or equivalently
	\begin{align}
	\label{eq:invertvelgen}
	\begin{bmatrix}
	\accentset{\ee}{\tilde{S}}_A{}^i\\
	\accentset{\ff}{\tilde{S}}_A{}^i
	\end{bmatrix}=\begin{bmatrix}
	\frac{\sqrt{\inducee}}{\lapsee}\accentset{\ee}{M}^{i\ j}_{\ A \ B} & \frac{\sqrt{\induceh}}{\lapseh}\Mkin^{i \ j}_{\ A\ B}\\
	\frac{\sqrt{\induceh}}{\lapseh}\Mkin^{i \ j}_{\ A \ B} & \frac{\sqrt{\inducef}}{\lapsef}\accentset{\ff}{M}^{i \ j}_{\ A \ B}
	\end{bmatrix}\begin{bmatrix}
	\dot{\ee}^B{}_j \\
	\dot{\ff}^B{}_j
	\end{bmatrix}\implies \begin{bmatrix}
	\dot{\ee}^B{}_j \\
	\dot{\ff}^B{}_j
	\end{bmatrix}=	\begin{bmatrix}
	\frac{\sqrt{\inducee}}{\lapsee}\accentset{\ee}{M}^{i\ j}_{\ A \ B} & \frac{\sqrt{\induceh}}{\lapseh}\Mkin^{i \ j}_{\ A\ B}\\
	\frac{\sqrt{\induceh}}{\lapseh}\Mkin^{i \ j}_{\ A \ B} & \frac{\sqrt{\inducef}}{\lapsef}\accentset{\ff}{M}^{i \ j}_{\ A \ B}
	\end{bmatrix}^{-1}\begin{bmatrix}
	\accentset{\ee}{\tilde{S}}_A{}^i\\
	\accentset{\ff}{\tilde{S}}_A{}^i
	\end{bmatrix}.
	\end{align}
	It is possible to find the Moore-Penrose pseudo inverse of a matrix to the block matrix in equation \eqref{eq:invertvelgen} analogous to \cite{Blixt:2018znp,Blixt:2019ene,Blixt:2019mkt}. This is done by using the usual inverse formula for block matrices, assuming that the determinants of the submatrices are non-zero. This can then be decomposed into irreducible parts under the rotation group, and if the corresponding primary constraints are imposed, then that part of the irreducible decomposition is simply removed in order to get the correct inverse. The Hamiltonian can then straightforwardly be derived. However, due to the lengthiness of these calculations we leave this for the future. Note that a proper 3+1 decomposition assumes that all tetrads are timelike with respect to each other and, hence, the following is true
	\begin{align}
	\normee_{A}\normhh^A\neq 0, \indent \inducee^{ij}\induceh_{ik}\neq 0, \indent \lapsee \neq 0, \indent \lapseh \neq 0, \indent \sqrt{\inducee} \neq 0, \indent \sqrt{\induceh} \neq 0,
	\end{align}
	and so on. This will also be useful when considering other bigravity theories.
	
	\subsection{Specific examples}
	\label{sec:Specificexamples}
	In this section we go into two examples for which we write down the Hamiltonian (to the step before using Dirac's algorithm to calculate Poisson brackets and find all constraints). The first example is two copies of new general relativity which is very easy to get from previous results \cite{Blixt:2018znp,Blixt:2019ene}. The second is the $\overset{kin}\LG$-term which share a lot of similarities with new general relativity. This theory, however, contains ghosts and only acts as a toy model. Note that $\HG\left(\overset{\ee}\LG+\overset{\ff}\LG+\overset{kin}\LG+\overset{int}\LG\right)\neq \overset{\ee}\HG+\overset{\ff}\HG+\overset{kin}\HG+\overset{int}\HG$. The inversion of velocities to conjugate momenta will in this case be different than simply the sum of them due to the derivative interactions between the tetrads in $\overset{kin}\LG$. The interaction potential transforms under Legendre transformation just by changing its sign. We, hence, focus only on kinetic terms for the Hamiltonian analysis (until the step when Poisson brackets are calculated).

	\subsubsection{Two copies of new general relativity: $\overset{\ee}\LG+\overset{\ff}\LG$}
	\label{sec:BiNGR}
	A straightforward example of a Hamiltonian for novel teleparallel bigravity terms is obtained by restricting to the case with no derivative interaction. Then the conjugate momenta for the tetrads will only be related to the velocity of the tetrad for which this conjugate momenta was defined. In this case, the Legendre transformation of the total Lagrangian equals the addition of the Lagrangians individually Legendre transformed,
	\begin{align}
	\HG\left(\overset{\ee}{\LG}+\overset{\ff}{\LG}\right)=\overset{\ee}{\HG}+\overset{\ff}{\HG}.
	\end{align}
	Since we already know $\overset{\ee}{\HG}$ from the literature \cite{Blixt:2018znp,Blixt:2019ene,Blixt:2019mkt}, and $\overset{\ff}{\HG}$ will just be another copy of the same Hamiltonian, but with respect to the $f$-tetrad we have
	\begin{align}\label{eq:twoNGRHam}
	\begin{split}
	\HG\left(\overset{\ee}{\LG}+\overset{\ff}{\LG}\right)&=\lapsee m_e^2\sqrt{\inducee}\left(-\frac{\CeVec_i\CeVec^i}{4A_\mathcal{V}}+\frac{\CeAn_{ij}\CeAn^{ij}}{4A_\mathcal{A}}+\frac{\CeSym_{ij}\CeSym^{ij}}{4A_\mathcal{S}}+\frac{3\CeTr\CeTr}{4A_\mathcal{T}}-\spatialee-\normee^A\covg_{i}\frac{\conje_A{}^i}{\sqrt{\inducee}} \right)\\
	&+\lapsef m_f^2\sqrt{\inducef}\left(-\frac{\CfVec_i\CfVec^i}{4\tilde{A}_\mathcal{V}}+\frac{\CfAn_{ij}\CfAn^{ij}}{4A_\mathcal{A}}+\frac{\CfSym_{ij}\CfSym^{ij}}{4A_\mathcal{S}}+\frac{3\CfTr\CfTr}{4A_\mathcal{T}}-\spatialff-\normff^A\covf_i\frac{\conjf_A{}^i}{\sqrt{\inducef}} \right)\\
	&+\shifte^{k}\sqrt{\inducee}\left(\torsionee^A{}_{jk}\frac{\conje_A{}^j}{\sqrt{\inducee}}-\ee^A{}_k\covg_i\frac{\conje_A{}^i}{\sqrt{\inducee}}\right)+\shiftf^{k}\sqrt{\inducef}\left(\torsionff^A{}_{jk}\frac{\conjf_A{}^j}{\sqrt{\inducef}}-\ff^A{}_k\covf_i\conjf_A{}^i\right)\\
	&+\sqrt{\inducee}\covg_i\left[\frac{\conje_A{}^i}{\sqrt{\inducee}}\left(\lapsee\normee^A+\shifte^j\ee^A{}_j\right)\right]+\sqrt{\inducef}\covf_i\left[\frac{\conjf_A{}^i}{\sqrt{\inducef}}\left(\lapsef\normff^A+\shiftf^j\ff^A{}_j\right)\right].
	\end{split}
	\end{align}
	Now the interaction potential should be added and it will modify the secondary constraints in lapse and shift. Since in general we do not have the special case where the antisymmetric modes become non-dynamical, we  avoid that non-linearities in lapse and shift appear from solving the equations of motion for the antisymmetric tetrad components. However, one should still be cautious since for particular theories there exists constraints making certain tetrad components non-dynamical and we might end up in a similar situation. From this point Poisson brackets need to be calculated for whichever sub-theory one is interested in, from which eventual secondary, tertiary, \dots constraints can be found. In particular, this theory has two subcases. One is new general relativity in the single tetrad limit. The other is teleparallel equivalent to Hassan-Rosen bigravity (see eq. \eqref{TeleBi}), when $\constonee=\constonef=\frac{1}{4}, \ \consttwoe=\consttwof=\frac{1}{2}, \ \mathrm{and} \ \constthreee=\constthreef=-1$. For any other choice of parameters we get novel theories of gravity. However, according to the linear analysis all those novel theories appear to be pathological.

	\subsubsection{Term with derivative interactions: $\overset{kin}\LG$}
	\label{sec:Thekinterm}
	We have already established the 3+1 decomposition of the Lagrangian in equation \eqref{eq:Lkin} and derived its primary constraints. In order to invert velocities into conjugate momenta we have to find the inverse of $M^{i \ j}_{ \ A \ B}$ which turns out to be
	\begin{align}\label{MInvIrr}
	\begin{split}
	\left(\tilde{M}^{-1}\right)^{A \ C}_{ \ i \ k}&=-\frac{1}{2}\kinveccoeffimposeconstrinv \normhh^A\normhh^C\induceh_{ik}+\frac{1}{2}\kinantcoeffimposeconstrinv\induceh^{r[s}\induceh^{m]n}\induceh_{kr}\induceh_{si}\tetradmean^A{}_{m}\tetradmean^C{}_n\\
	&+\frac{1}{2}\kinsymcoeffimposeconstrinv\left(\induceh^{r(s}\induceh^{m)n}-\frac{1}{3}\induceh^{sm}\induceh^{nr}\right)\induceh_{kr}\induceh_{si}\tetradmean^A{}_m\tetradmean^C{}_n+\frac{1}{6}\kintrcoeffimposeconstrinv\tetradmean^A{}_{i}\tetradmean^C{}_k,
	\end{split}
	\end{align}
	where the irreducible parts are identified using $(\mathcal{I} \ \in \ \{\mathcal{V},\mathcal{A},\mathcal{S},\mathcal{T}\})$ $\kincoeffimposeconstrinv=\begin{cases}
	0 \ \mathrm{for} \ \kincoeffimposeconstr =0\\
	\frac{1}{\kincoeffimposeconstr} \ \mathrm{for} \ \kincoeffimposeconstr \neq 0
	\end{cases}$.
	We can write the Hamiltonian as
	\begin{align}
	\begin{split}
	\label{HwithMinv}
	\HG&=\lapseh\Bigg[ \frac{\conje_A{}^i\conjf_B{}^j}{m_{ef}^2\sqrt{\induceh}}\left(\tilde{M}^{-1}\right)^{A \ B}_{\ i\ j}-\left(\conje_B{}^j\accentset{\ee}{T}^C{}_{kl}+\conjf_B{}^j\accentset{\ff}{T}^C{}_{kl}\right)\induceh^{il}\left(\consttwokin\normhh_C\cotetradmean_A{}^k+\constthreekin\normhh_A\cotetradmean_C{}^k\right)\left(\tilde{M}^{-1}\right)^{A \ B}_{\ i \ j}
	\\&+m_{ef}^2\sqrt{\induceh}\accentset{\ee}{T}^C{}_{kl}\accentset{\ff}{T}^D{}_{mp}\induceh^{ip}\induceh^{jl}\left(\consttwokin\normhh_D\cotetradmean_A{}^m+\constthreekin\normhh_A\cotetradmean_D{}^m\right)\left(\consttwokin\normhh_C\cotetradmean_B{}^k+\constthreekin\normhh_B\cotetradmean_C{}^k\right)\left(\tilde{M}^{-1}\right)^{A \ B}_{\ i \ j}-m_{ef}^2\sqrt{\induceh}\cdot \spatialhh\Bigg]
	\\&+\shifth^i\left[\conje_B{}^j\accentset{\ee}{T}^B{}_{ij}+\conjf_B{}^j\accentset{\ff}{T}^B{}_{ij}\right]+\sqrt{\inducee}\covg_i\ee^A{}_0\frac{\conje_A{}^i}{\sqrt{\inducee}}+\sqrt{\inducef}\covf_i\ff^A{}_0\frac{\conjf_A{}^i}{\sqrt{\inducef}}.
	\end{split}
	\end{align}
	We have that
	\begin{align}
	\begin{split}
	\label{SymmetrieswithMinv}
	\left[\consttwokin\induceh^{il}\normhh_B\cotetradmean_A{}^k+\constthreekin\induceh^{il}\normhh_A\cotetradmean_B{}^k\right]\left(\tilde{M}^{-1}\right)^{\ A \ C}_{i \ m}&=- \kinveccoeffimposeconstrinv \constthreekin\delta^l_m\normhh^C\cotetradmean_B{}^k+\kinantcoeffimposeconstrinv\left(-\frac{1}{2}\consttwokin\normhh_B\cotetradmean^{Ck}\delta^l_m+\frac{1}{2}\consttwokin\normhh_B\cotetradmean^{Cl}\delta^k_m\right)
	\\&+\kinsymcoeffimposeconstrinv\left(-\frac{1}{2}\consttwokin\normhh_B\cotetradmean^{Ck}\delta^l_m+\frac{1}{3}\consttwokin\induceh^{kl}\normhh_B\cotetradmean^C{}_m-\frac{1}{2}\consttwokin\normhh_B\cotetradmean^{Cl}\delta^k_m\right)
	\\&-\frac{\kintrcoeffimposeconstrinv}{3}\consttwokin\induceh^{kl}\normhh_B\cotetradmean^C{}_m.
	\end{split}
	\end{align}
	We note that this expression appears in \eqref{HwithMinv} and is always contracted with something antisymmetric in $k$ and $l$. One can now quite easily show, by the use of the decomposition of the conjugate momenta into irreducible parts \eqref{eq:irrconjE} and \eqref{eq:irrconjF}, that the Hamiltonian expressed in the possible primary constraints becomes:
	\begin{align}
	\begin{split}
	\HG&=\lapseh m_{ef}^2\sqrt{\induceh}\Bigg[ \frac{\kinveccoeffimposeconstrinv}{2} {}^\mathcal{V}C^i{}^\mathcal{V}\tilde{C}^j\induceh_{ij}-\frac{\kinantcoeffimposeconstrinv}{2}{}^\mathcal{A}C^{ij}{}^\mathcal{A}\tilde{C}^{kl}\induceh_{ik}\induceh_{jl}-\frac{\kinsymcoeffimposeconstrinv}{2}{}^\mathcal{S}C^{ij}{}^\mathcal{S}\tilde{C}^{kl}\induceh_{ik}\induceh_{jl}-\frac{3\kintrcoeffimposeconstrinv}{2}{}^\mathcal{T}C{}^\mathcal{T}\tilde{C}-\spatialhh  \Bigg]
	\\ & +\shifth^i\left[\conje_B{}^j\accentset{\ee}{T}^B{}_{ij}+\conjf_B{}^j\accentset{\ff}{T}^B{}_{ij}\right]+\partial_i\ee^A{}_0\conje_A{}^i+\partial_i\ff^A{}_0\conjf_A{}^i.
	\end{split}
	\end{align}
	Now we perform an integration by parts and get:
	\begin{align}
	\begin{split}
	\HG&=\lapseh m_{ef}^2\sqrt{\induceh}\Bigg[ \frac{\kinveccoeffimposeconstrinv}{2} {}^\mathcal{V}C^i{}^\mathcal{V}\tilde{C}^j\induceh_{ij}-\frac{\kinantcoeffimposeconstrinv}{2}{}^\mathcal{A}C^{ij}{}^\mathcal{A}\tilde{C}^{kl}\induceh_{ik}\induceh_{jl}-\frac{\kinsymcoeffimposeconstrinv}{2}{}^\mathcal{S}C^{ij}{}^\mathcal{S}\tilde{C}^{kl}\induceh_{ik}\induceh_{jl}-\frac{3\kintrcoeffimposeconstrinv}{2}{}^\mathcal{T}C{}^\mathcal{T}\tilde{C}-\spatialhh  \Bigg]
	\\ & +\shifth^i\left[\conje_B{}^j\accentset{\ee}{T}^B{}_{ij}+\conjf_B{}^j\accentset{\ff}{T}^B{}_{ij}\right]-\sqrt{\inducee}\left(\lapsee\normee^A+\shifte^j\ee^A{}_j\right)\covg_i\frac{\conje_A{}^i}{\sqrt{\inducee}}-\left(\lapsef\normff^A+\sqrt{\inducef}\shiftf^j\ff^A{}_j\right)\covf_i\frac{\conjf_A{}^i}{\sqrt{\inducef}}
	\\&+\sqrt{\inducee}\covg_i\left(\ee^A{}_0\frac{\conje_A{}^i}{\sqrt{\inducee}}\right)+\sqrt{\inducef}\covf_i\left(\ff^A{}_0\frac{\conjf_A{}^i}{\sqrt{\inducef}}\right).
	\end{split}
	\end{align}
	Again, the fact that the equations of motion for the antisymmetric parts of the tetrads introduce non-linearities in lapse and shift is avoided. One still needs to be cautious for certain choices of parameters that one can still end up in similar situations. Now it is time to choose the metric $\gh$. One prominent choice would be the so-called geometric mean \cite{Kocic:2018ddp,Kocic:2018yvr} for which the two metrics $\gee$ and $\gf$ appear on equal footing by satisfying \eqref{eq:geometricmean}. However, this could only be viewed as a toy model when demanding the theory to be ghost-free. Another prominent choice would be to consider $\gh$ constructed from complex tetrads similar to what have been considered in \cite{Apolo:2016ort,Apolo:2016vkn}. In this case more primary constraints are expected in the Hamiltonian analysis and the shape of the Hamiltonian may alter. For the case of the linear analysis it would be considerably different involving complex tetrads and this is, thus, beyond the scope of this article.

	\section{Discussion and conclusions}
	The teleparallel equivalent of Hassan-Rosen bimetric gravity (TEHR) has been formulated and it admits almost the same structure of constraints as the tetrad formulation of Hassan-Rosen bimetric gravity. The primary constraints associated with the antisymmetric part of the conjugate momenta differ, however, by the presence of torsion terms originating from differences in the boundary \cite{Blixt:2020ekl}. Nevertheless, the symmetries the theory admits are the same, and thus, the degrees of freedom coincide with other formulations of Hassan-Rosen bimetric gravity \cite{Hassan:2011zd,Hinterbichler:2012cn}. With TEHR as a starting point we found several possibilities to create novel modified theories of gravity. For instance, one can make two copies of NGR. As an initial assumption, the NGR coefficients in the $\gee$ sector can be different from the $\gf$ sector. However, analyzing the perturbations around the Minkowski background reveals that the coefficients need to be proportional to each other for the modes to be decoupled. The only difference is absorbed in the mass constant in front of the respective kinetic term used to define the Fierz-Pauli mass. It is already known that NGR suffer from the strong coupling problem in Minkowski backgrounds \cite{Cheng:1988zg,BeltranJimenez:2019nns} and ghost instabilities in the generic case \cite{Ortin:2015hya}. We confirm this result and find that the problem worsens in the case of teleparallel bigravity since we find that ghostly modes propagate in this theory\footnote{The strong coupling problem could in theory be avoided by going to less trivial backgrounds and analyze the viability of the theory in more generic backgrounds. An example of such an attempt can be found in $f(T)$-gravity \cite{Golovnev:2020nln}.}.

	As another example, we introduce terms with derivative interactions. It appears that the teleparallel framework makes construction of such examples more easy, however, likewise the Riemannian case is known to introduce ghost instabilities in the symmetric sector \cite{Boulanger:2000rq} in the massless case, this may be retained in the massive case. In future works, it would be interesting to investigate constructions with complex tetrads similar to what have been done in the Riemannian case \cite{Apolo:2016ort,Apolo:2016vkn}. The motivation for this would be to investigate if viable theories with partially massless symmetry can be constructed in a viable way. Previous results suggest that this is strongly coupled around Minkowski backgrounds, and one would need to extend the linear analysis in this work to perturbations around complex De Sitter tetrads. It is also outlined how the Hamiltonian can be obtained in the model considered in this article, and with more details in a simpler model where the derivation is easier to carry out.

	Furthermore, it is shown in app. \ref{appendix:GaugeFixing} that as in other teleparallel theories \cite{Blixt:2018znp,Blixt:2019mkt,Golovnev:2021omn,Blixt:2022rpl,Golovnev:2023yla}, the so-called Weitzenb{\"o}ck gauge can safely be assumed. One may easily be mistaken, trying to include two spin connections in the construction of teleparallel bigravity theories since bimetric gravity assumes two Levi-Civita connections. However, that would have little sense and miss the point of the theory covariant under Lorentz transformations \cite{Golovnev:2017dox,Krssak:2015oua}. Overall, our work has made the initial investigations for teleparallel bigravity. Providing details about teleparallel and bimetric gravity and explained the difference between these theories and a couple of other theories that would appear to be similar. Some novel theories have been ruled out by our analysis, which also provides useful tools to further investigate perturbation theory and Hamiltonian analysis in the future.
	
	Currently, the perhaps most promising avenue to pursue is paved by the parameterised field theory approach to gauging translations \cite{Koivisto:2019ejt,Koivisto:2022uvd}. The conventional approach, adopted also in this paper, is well established to describe gauge interactions of fields and particles {\it in} spacetime, in continuum as well as in lattice models. However, if we would rather consider gravity as the gauge theory {\it of} spacetime, we should not assume the inhomogeneous extension of the symmetry group to begin with. In the parameterised field theory approach, the fundamental field giving rise to the spacetime metric is not its extension to a tetrad, but its reduction to a scalar. Recently, a viable theory with two interacting frame fields emerging from two scalar fields and giving rise to two independent but non-trivially coupled metrics, has been constructed in this approach, though not for bigravity but for the unification of gravity and the particle physics Yang-Mills theories \cite{Gallagher:2022kvv}. Some of the methods and results we presented in this paper could be useful also in the potentially interesting application of the parameterised field theory approach to interacting multiple spin-2 field theories. 
	
	\section*{Acknowledgments}
The authors are grateful to Fawad Hassan for very interesting discussions.
MH LM and TK gratefully acknowledge the full financial support by the Estonian Research Council through the Personal Research Funding project PRG356 and by the European Regional Development Fund through the Center of Excellence TK133 ``The Dark Side of the Universe''.

	\appendix
	\section{Gauge fixing}
	\label{appendix:GaugeFixing}
	We show here that the Hamiltonian analysis can be done in the so-called Weitzenb{\"o}ck gauge in all theories considered in this article. In \cite{Blixt:2019mkt,Blixt:2018znp,Blixt:2022rpl,Golovnev:2021omn,Golovnev:2023yla} it was shown that this is valid in a quite general class of teleparallel theories. However, the results do not cover teleparallel bigravity theories. In this appendix, we will conduct a quite different notation, where we define conjugate momenta without a priori gauge fix, so that the spin connection $\omega^A{}_{B\mu}=0$. Furthermore, we denote Lorentz transformations by ``\ \~\ ''. Consider the teleparallel bigravity theory given by \eqref{GenBiGrav}. In this expression we know that all derivatives appear in $\accentset{\ee}{T}^A{}_{\mu\nu}$ and $\accentset{\ff}{T}^A{}_{\mu\nu}$. In order to account for the symmetries of the spin connection, we introduce the same auxiliary field as was introduced in \cite{Blixt:2019mkt,Blixt:2018znp}
	\begin{align}
	\aux_{AB}:=\eta_{AC}\omega^C{}_{B0}=\eta_{C[A}\Lambda^C{}_{|D|}\partial_0 \left(\Lambda^{-1}\right)^D{}_{B]}\Leftrightarrow \partial_0 \Lambda^A{}_B=a_{CD}\eta^{A[D}\Lambda^{C]}{}_B.
	\end{align}
	We introduce conjugate momenta with respect to both tetrads
	\begin{align}
	\conje_A{}^i=\frac{\partial \LG}{\partial \partial_0 \ee^A{}_i}, \indent \conjf_A{}^i=\frac{\partial \LG}{\partial \partial_0 \ff^A{}_i}.
	\end{align}
	The conjugate momenta with respect to Lorentz matrices are inherited by the following expression
	\begin{align}
	\conjspin^{AB}:=\frac{\partial \LG}{\partial \aux_{AB}},
	\end{align}
	since the auxiliary field $\aux_{AB}$ is linearly related to the time derivatives of the Lorentz matrices. Explicitly, we have
	\begin{align}
	\conje_A{}^i=2\accentset{\ee}{T}^B{}_{\rho \sigma}\accentset{\ee}{G}_{AB}{}^{0i\rho\sigma}+\accentset{\ff}{T}^B{}_{\rho \sigma}\accentset{kin}{G}_{AB}{}^{0i\rho\sigma},\\
	\conjf_A{}^i=\accentset{\ee}{T}^B{}_{\rho\sigma}\accentset{kin}{G}_{BA}{}^{\rho\sigma 0i}+2\accentset{\ff}{T}^B{}_{\rho \sigma}\accentset{\ff}{G}_{AB}{}^{0i\rho\sigma},\\
	\conjspin^{AB}=\frac{\partial \LG}{\partial\accentset{\ee}{T}^C{}_{0i}}\frac{\partial\accentset{\ee}{T}^C{}_{0i}}{\partial a_{AB}}+\frac{\partial \LG}{\partial\accentset{\ff}{T}^C{}_{0i}}\frac{\partial\accentset{\ff}{T}^C{}_{0i}}{\partial a_{AB}}=\conje_C{}^i\frac{\partial\accentset{\ee}{T}^C{}_{0i}}{\partial a_{AB}}+\conjf_C{}^i\frac{\partial\accentset{\ff}{T}^C{}_{0i}}{\partial a_{AB}}.
	\end{align}
	Furthermore, we have
	\begin{align}
	\begin{split}
	\frac{\partial\accentset{\ee}{T}^C{}_{0i}}{\partial a_{AB}}=-\eta^{C[B}\ee^{A]}{}_i,	\\ \frac{\partial\accentset{\ff}{T}^C{}_{0i}}{\partial a_{AB}}=-\eta^{C[B}\ff^{A]}{}_i.
	\end{split}
	\end{align}
	This means that we have the following algebraic relation
	\begin{align}\label{eq:WeitzenboeckConstraint}
	\conjspin^{AB}=-\conje_C{}^i\eta^{C[B}\ee^{A]}{}_i-\conjf_C{}^i\eta^{C[B}\ff^{A]}{}_i,
	\end{align}
	for which we get the following primary constraints
	\begin{align}
	\label{eq:LorPC}
	{}^{\pi}\lambda^{AB}\equiv \conjspin^{AB}+\conje_C{}^i\eta^{C[B}\ee^{A]}{}_i+\conjf_C{}^i\eta^{C[B}\ff^{A]}{}_i.
	\end{align}
	Under Lorentz transformation (denoted with ``\ \~ \ ''), the conjugate momenta transforms as
	\begin{align}
	\tilde{\conje}_A{}^i&=\pi_B{}^i\Lambda^B{}_A,\\
	\tilde{\conjf}_A{}^i&=\conjf_B{}^i\Lambda^B{}_A,\\ \label{eq:LorentzMomenta}
	\tilde{\conjspin}^{MN}&=\conje_A{}^j\eta^{A[N}\ee^{M]}{}_j+\conjf_A{}^{j}\eta^{A[N}\ff^{M]}{}_j+\conjspin^{MN},
	\end{align}
	with the inverse transformation
	\begin{align}
	\conje_A{}^i&=\tilde{\conje}_B{}^i\left(\Lambda^{-1}\right)^B{}_A,\\
	\conjf_A{}^i&=\tilde{\conjf}_B{}^i\left(\Lambda^{-1}\right)^B{}_A,\\
	\conjspin^{MN} &=\tilde{\conjspin}^{MN}-\conje_A{}^j\eta^{A[N}\ee^{M]}{}_j-\conjf_A{}^j\eta^{A[N}\ff^{M]}{}_j
	\end{align}
	By imposing the primary constraint, given by eq. \eqref{eq:WeitzenboeckConstraint}, to eq. \eqref{eq:LorentzMomenta} we see that $\tilde{\conjspin}^{MN}\approx 0$. We can now look at the Lorentz transformed Hamiltonian which schematically will look like
	\begin{align}
	\begin{split}
	\tilde{\mathcal{H}}\left[{}^{\tilde{\pi}}\lambda,\tilde{\conje},\tilde{\conjf},
	\tilde{\conjspin},\ee,\ff,\tilde{\Lambda}\right]&=\tilde{\conje}_A{}^i\tilde{\dot{\ee}}^A{}_i+\tilde{\conjf}_A{}^i\tilde{\dot{\ff}}^A{}_i+\tilde{\conjspin}^{AB}\tilde{a}_{AB}+\tilde{{}^{\pi}}\lambda_{AB}\tilde{\conjspin}^{AB}-\tilde{\LG}\left[\tilde{\ee},\tilde{\dot{\ee}},\tilde{\ff},\tilde{\dot{\ff}},\tilde{\Lambda}\right]+\mathrm{P.C},
	\end{split}
	\end{align}
	where $\mathrm{P.C}$ are theory specific primary constraints. It turns out that $\tilde{\LG}\left[\tilde{\ee},\tilde{\dot{\ee}},\tilde{\ff},\tilde{\dot{\ff}},\tilde{\Lambda}\right]=\tilde{\LG}\left[\tilde{\ee},\tilde{\dot{\ee}},\tilde{\ff},\tilde{\dot{\ff}}\right]$. Furthermore, the Hamiltonian is on-shell independent of both $\tilde{\Lambda}$ and $\tilde{\conjspin}$. The evolution of the primary constraints eq. \eqref{eq:LorPC} is expected to vanish on-shell similar to the case of \cite{Golovnev:2021omn}. Hence, the number of degrees of freedom obtained by the Hamiltonian analysis of \eqref{GenBiGrav} in the Weitzenb{\"o}ck gauge will not differ from the canonical Hamiltonian analysis. The results of \cite{Blixt:2022rpl,Golovnev:2021omn,Golovnev:2023yla} can easily be extended to the case of teleparallel bigravity as well. The important point is that the covariantization only involves one spin connection. This is the only sensible way to make the theory Lorentz covariant similar to the more studied cases \cite{Golovnev:2017dox}. The same conclusions about the consistency of working in the Weitzenb{\"o}ck gauge from the start can be argued in a similar way as was done elsewhere in the literature \cite{Blixt:2022rpl,Golovnev:2023yla}.

\providecommand{\href}[2]{#2}\begingroup\raggedright\endgroup

\end{document}